\documentclass[12pt,preprint]{aastex}
\usepackage{emulateapj5}
\usepackage{apjfonts}
\usepackage{graphicx}

\newcommand {\apgt} {\ {\raise-.5ex\hbox{$\buildrel>\over\sim$}}\ }
\newcommand {\aplt} {\ {\raise-.5ex\hbox{$\buildrel<\over\sim$}}\ }
\usepackage{natbib}
\usepackage{graphicx}
\usepackage{apjfonts}
\usepackage{color}

\slugcomment{Published in The Astrophysical Journal Supplement Series, 189:1--14, 2010 July}

\shorttitle{A 3.5\,mm Polarimetric Survey of Radio-loud Active Galactic Nuclei}
\shortauthors{Agudo et al.}

\begin{document}

\title{A 3.5\,mm Polarimetric Survey of Radio-loud Active Galactic Nuclei}

\author{I. Agudo\altaffilmark{1, 2}, C. Thum\altaffilmark{3}, 
            H. Wiesemeyer\altaffilmark{4}, T.~P. Krichbaum\altaffilmark{5}}

\altaffiltext{1}{Instituto de Astrof\'{\i}sica de Andaluc\'{\i}a (CSIC),  Apartado 3004, E-18080 Granada, Spain}

\altaffiltext{2}{Institute for Astrophysical Research, Boston University, 725 Commonwealth Avenue, Boston, MA 02215; \email{iagudo@bu.edu}}

\altaffiltext{3}{Institut de Radio Astronomie Millim\'etrique, 300 Rue de la Piscine, 38406 St. Martin d'H\`eres, France;  \email{thum@iram.fr}}

\altaffiltext{4}{Instituto de Radio Astronom\'{\i}a Milim\'etrica, Avenida Divina Pastora, 7, Local 20, E-18012 Granada, Spain; \email{wiesemey@iram.es}}                

\altaffiltext{5}{Max-Planck-Institut f\"ur Radioastronomie, Auf dem H\"ugel, 69, D-53121 Bonn, Germany; \email{tkrichbaum@mpifr-bonn.mpg.de}}  

\begin{abstract}
      We present the results from the first large ($>100$ source) 3.5\,mm polarimetric survey of radio loud active galactic nuclei (AGN).
      This wavelength is favorable within the radio--mm range for measuring the intrinsic linearly polarized emission from AGN, since in general it is only marginally affected by Faraday rotation of the electric vector position angle, and depolarization.
      The  $I$, $Q$, $U$, and $V$ Stokes parameter observations were performed with the XPOL polarimeter at the IRAM 30\,m Telescope on different observing epochs from July 2005 (when most of the measurements were made) to October 2009.
     Our sample consists of 145 flat-radio-spectrum AGN with declination $>-30^{\circ}$ (J2000.0) and flux density $\apgt1$\,Jy at $\sim86$\,GHz, as measured at the IRAM 30\,m Telescope from 1978 to 1994. 
     This constraint on the radio spectrum causes our sample to be dominated by blazars, which allows us to conduct new statistical studies on this class of high-luminosity, relativistically-beamed emitters.
     We detect linear and circular polarization (above minimum $3\sigma$ levels of $\sim$1.5\,\%, and $\sim$0.3\,\%) for 76\,\%, and 6\,\% of the sample, respectively.
     We find a clear excess in degree of linear polarization detected at 86\,GHz with regard to that at 15\,GHz by a factor of $\sim2$. 
     Over our entire source sample, the luminosity of the jets is anti--correlated with the degree of linear polarization.
     Consistent with previous findings claiming larger Doppler factors for brighter $\gamma$-ray blazars, quasars listed in our sample, and in the \emph{Fermi} Large Area Telescope Bright Source Catalog (LBAS), show larger luminosities than non--LBAS ones, but our data do not allow us to confirm the same for BL~Lac objects.
     We do not find a clear relation between the linear polarization angle and the jet structural position angle for any source class in our sample. We interpret this as the consequence of a markedly non-axisymmetric character of the 3\,mm emitting region in the jets.
     We find that intrinsic circular polarization is the most likely mechanism for generation of the circular polarization detected in our observations.
     Our new data  can be used to estimate the 3.5\,mm AGN contribution to measurements of the linear polarization of the cosmic microwave background, such as those performed by the Planck satellite. 
\end{abstract}


\section{Introduction}
\label{Intr}

Radio loud active galactic nuclei (AGN) are known to produce powerful pairs of highly collimated relativistic jets of magnetized plasma, which are able to extend far beyond the boundaries of the host galaxy. 
Their relativistic nature, their magnetic fields, and their non-isotropic geometry determine their most relevant observational properties: superluminal motions \cite[e.g.,][]{Gomez:2001p201, Jorstad:2005p264}, Doppler boosted (decreased) emission of the jet pointing at a small (large) angle to the observer \citep[e.g.,][]{Kadler:2004p5629}, rapid intrinsic emission variability \cite[e.g.,][]{Agudo:2006p203, Fuhrmann:2008p267, 2008Natur.452..966M} and changes of the jet structure \cite[e.g.,][]{2007AJ.134.799J, Agudo:2007p132}, and intense polarized synchrotron and inverse-Compton emission along all the electromagnetic spectrum \cite[e.g.,][]{Ostorero:2006p406, Abdo:2010, Marscher:2010p11374}.

Polarimetric very long baseline interferometry (VLBI) observations at mm wavelengths, combined with single dish radio, mm, and optical linear polarimetric observing campaigns, allow one to (i) connect the location of the emitting regions at different observing bands and (ii) infer properties about the nature of the innermost moving and stationary knots of emission in the jets and the magnetic field in such regions, which usually cannot be resolved by VLBI \cite[e.g.,][]{2007AJ.134.799J, 2008Natur.452..966M, Marscher:2010p11374}.

VLBI polarimetric surveys of radio loud AGN also provide relevant information about their jets; see \citet{2003ApJ...589..733P} for a survey on a sample of 177 sources in the Calteck-Jodrell Bank Flat-Spectrum survey observed with the Very Long Baseline Array (VLBA) at 5\,GHz, and \citet{Lister:2005p261, Homan:2006p238} for the first results from the MOJAVE survey on 133 sources observed with the VLBA at 15\,GHz.
\citet{Lister:2005p261} and \citet{2003ApJ...589..733P} found the source cores (i.e., the innermost visible jet region at a given observing frequency with an instrument capable of resolving such a jet) to be weakly linearly polarized (\aplt5\,\%), but with significantly larger fractional polarization for the jet regions downstream.
Indeed, \citet{Lister:2005p261} reported a general increase of linear polarization degree with increasing distance outward from the core in quasars and BL~Lacertae (BL~Lac) objets, but with BL~Lac jets more polarized in general.
They also found the cores and jets of radio galaxies to be much weaker or not linearly polarized compared to those in quasars and BL~Lac objects. 
\citet{Homan:2006p238} found ``strong'' circular polarization ($\ge$0.3\%, as defined by them), usually in the cores of the sources (but not always), within $\sim$15\% of their sample. 
They did not find significant correlations between the degree of circular polarization in the core and other relevant source properties. 

No polarization survey of a large number of AGN has been performed at mm wavelengths thus far.  
It is known that AGN jets and their cores are affected by non-negligible Faraday rotation measures (RM) typically ranging from $\sim10^2$\,rad\,m$^{-2}$ \citep[e.g.,][]{2003ApJ...589..126Z, Zavala:2004p138, Gabuzda:2001p378, 2004MNRAS.351L..89G, 2008ApJ...675...79A} to $\sim10^4$\,rad\,m$^{-2}$ \citep[e.g.,][]{2002ApJ...566L...9Z, Attridge:2005p220, Gomez:2008p30}.
These large RM values $\sim10^4$\,rad\,m$^{-2}$ have only been found for a few sources, typically in the jets of radio galaxies  \citep{Zavala:2004p138}.
Since the observed electric vector linear polarization angle of a source $\chi_{\rm{obs}}=\chi_{\rm{int}}+\rm{RM}~\lambda^{2}$ (where $\rm{RM}~\lambda^{2}$ is the amount of Faraday rotation and $\chi_{\rm{int}}$ is the intrinsic polarization angle), a $\lambda=$3.5\,mm (86\,GHz) survey would be much less affected by Faraday rotation than are cm wavelength observations. 
At this wavelength, $|\rm{RM}|\approx7000$\,rad\,m$^{-2}$ would be required to produce Faraday rotation by $\approx5$\,$^{\circ}$, the median linear polarization angle uncertainty in the measurements presented in this paper.
In contrast, such a value of $|\rm{RM}|$ translates into rotations of $\approx160^{\circ}$ and $\approx1400^{\circ}$ at 2\,cm (15\,GHz) and 6\,cm (5\,GHz), respectively.
If AGN jet RM in the radio and mm emitting regions are similar in general -- which is still to be confirmed observationally-- the reduced effect of Faraday rotation in mm wavelength observations also makes them much less sensitive to Faraday depolarization, hence reflecting the intrinsic linear polarization properties of the sources with better fidelity than radio observations.

The mm wavelength emission of radio loud AGN is typically dominated by compact regions in their jets; either in their innermost cores of emission, or in their pc scale superluminal knots \citep[e.g.,][]{2007AJ.134.799J, 2008Natur.452..966M}. 
Indeed, the results from the 86\,GHz VLBI survey of \citet{Lee:2008p301} have shown that, in general, $\sim50$\,\% of the total flux density of their imaged sources is concentrated in regions not further than $\sim1.5$\,milli-arcseconds from the mm core.  
In contrast, because of the longer lifetime of synchrotron radiation at cm wavelengths \citep{Rybicki:1979p6159}, such emission is more spread out along the pc scale of the jets up to distances $>5$\,milli-arcseconds \citep{2003ApJ...589..733P, Lister:2005p261}.
Moreover, owing to synchrotron opacity, the bright cm wavelength emission from compact cores in AGN lies in regions farther downstream in the jet than at mm wavelengths \citep[e.g.,][]{Lobanov:1998p5354, Marscher:2006p362, 2007AJ.134.799J}.
 
In the absence of a large mm-VLBI polarimetric survey, the statistical analysis of a single-dish polarization-mm survey can yield relevant insights into the intrinsic polarization properties of the innermost compact regions of relativistic jets in AGN.
Here we present the results from the first 3.5\,mm polarimetric survey over a large ($>100$ source) sample of radio loud AGN.

Our results are also a valuable resource for surveys dedicated to studies of the polarization properties of the cosmic microwave background (CMB), specially that conducted by the ESA's Planck mission (see the mission \emph{Bluebook}\footnote{\tt www.rssd.esa.int/SA/PLANCK/docs/Bluebook-ESA-SCI(2005)1\_V2.pdf}).
For such studies, radio loud AGN are considered the most important CMB-foreground contributors away from the Galactic plane at angular scales $\aplt0.5^{\circ}$ up to $\sim100$\,GHz observing frequency \citep{Tucci:2005p8610}.
Hence, \emph{a priori} knowledge of their polarization is important to account for their contribution to the polarization of the CMB data \citep[e.g.,][]{RubinoMartin:2008p8597,LopezCaniego:2009p8509}.  

\section{The Sample}
\label{Samp}

\begin{figure}
   \centering
   \includegraphics[width=8.5cm]{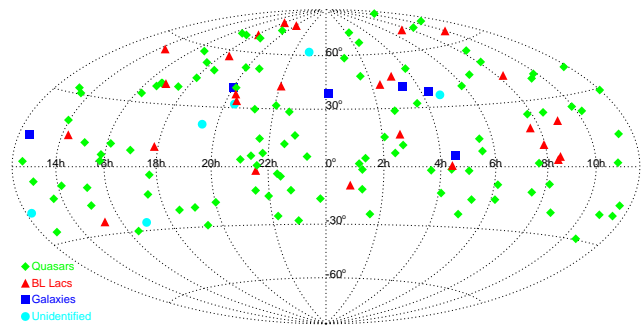}
   \caption{Sky distribution of the source sample in J2000.0 equatorial coordinates.}
   \label{skymap}
\end{figure}

In Table~\ref{T1} we present our sample of 145 observed sources (see also Fig.~\ref{skymap}).
This sample has been built essentially from the 138 radio loud AGN in the IRAM pointing source list.
Such list comes from the sample of $\sim300$ radio-to-mm flat-spectrum compact AGN visible from the IRAM 30\,m Telescope selected by \citet{Steppe:1988p2440} that were contained in the 1\,Jy catalog \citep{Kuehr:1981p6040} and/or in the catalog of positions, structures, and polarizations of 404 compact radio sources by \citet{Perley:1982p6054}.

From this sample, \citet{Steppe:1992p2441, Steppe:1993p2453} and \citet{Reuter:1997p2467} selected the brightest 138 AGN sources measured by the IRAM 30\,m AGN monitoring program with a typical flux density $S_{90}\apgt1$\,Jy at $\sim90$\,GHz and a minimum at $S_{90}\sim0.5$\,Jy in the time span from 1978 to 1994.
After removing the 4 sources with J2000.0 declination $<-30^{\circ}$ (too low to be observed at the 30\,m Telescope above $20^{\circ}$ elevation), we complemented our sample with 11 new sources not contained in the IRAM pointing source catalog but strong enough currently to be considered within the selection criteria for such a catalog.

Since the sample by \citet{Steppe:1992p2441, Steppe:1993p2453} and \citet{Reuter:1997p2467}  was essentially defined to be flux density limited at $S_{90}\apgt$1\,Jy from 1978 to 1994, our sample was expected to be complete in that sense. Most likely because of source variability,  54 (19) sources in our sample have measured flux densities $S_{86}<$1\,Jy ($S_{86}<$0.5\,Jy), hence not matching, at the observing epoch, the constraints imposed for the 1978 to 1994 time span.

To have a raw estimate of the number of radio loud AGN with $S_{86}\apgt$1\,Jy that we may have missed in our sample, we compared it with the 5--Year WMAP Point Source Catalog \citep[WMAP5,][]{Wright:2009p6229}.
From this catalog, we find 28 sources with J2000.0 declination $\ge-30^{\circ}$, and with flux density larger than 1\,Jy at 90\,GHz but not contained in our sample.
In contrast, our sample contains 20 sources with $S_{86}>$1\,Jy that are not found in WMAP5. 
All but one of these are low Galactic latitude sources, which reflects the bias of the WMAP5 catalog that excludes sources near the Galactic plane.

Assuming that a 1\,Jy flux--limited complete sample of radio loud AGN above $-30^{\circ}$ dec. contains all sources with $S_{86}>$1\,Jy both in our sample (i.e. 91), and in WMAP5, we only miss 24\,\% of eligible sources. 
Thus, although this does not allow us to claim completeness of our sample to the 1\,Jy total flux density limit, the relatively small fraction of missing sources, together with the unprecedented large size of our 86 GHz polarization sample, allows us to rely on the results derived from it as long as subsample selection does not lead to unacceptably small number statistics, which is not the case for any of our subsamples (see below).

In a survey like ours, polarization non-detection is a relevant bias that may be introduced by source variability when the total flux density of a considerable fraction of the sample decreases.
However, we have checked that we are not very much affected by such bias.
Among our 91 sources with $S_{86}\ge$1\,Jy, 77\,\% were detected in linear polarization, whereas for the 54 sources with $S_{86}<$1\,Jy, the fraction of linear polarization detections is similar (74\,\%).

Given the spectral criteria applied to select the sources for the IRAM pointing source list, our sample is dominated by radio-to-mm flat-spectrum compact AGN, i.e., by blazars.
To be more specific, it contains 107 quasars,  26 BL~Lac objects, and  6 radio galaxies, with 6 unclassified sources, i.e., not contained in the \citet{VeronCetty:2006p4900} catalog.
The sample also contains 32 sources in the \emph{Large Area Telescope} \citep[LAT,][]{Atwood:2009p6950}  $\gamma$-ray bright AGN source sample (LBAS).
The LBAS list contains sources detected by LAT onboard the \emph{Fermi Gamma-ray Space Telescope} during its first 3 months of operation   \citep[i.e., those in Tables~1 and 2 of][]{Abdo:2009p7779}.
These 32 LBAS sources include 19 quasars,  12 BL~Lacs, and, 1 radio galaxy. 
Note that out of the 117 LBAS sources in \citet{Abdo:2009p7779}, only 97 are visible from the 30\,m Telescope with declination $>-30^{\circ}$. 
Among these 97, 32 (33\,\%) are in our 3.5\,mm sample. 
The source redshift in the sample ranges from $z=0.004$ to $z=3.408$, with mean and median  $\bar{z}=0.941$ and $\tilde{z}=0.815$, respectively.

Hence, our sample is adequate for studies of the mm polarimetric properties of quasar and BL~Lac blazars, and of their relation with their  $\gamma$-ray properties.
Both the MOJAVE and our samples were selected primarily based on the flatness of source radio spectra.
Hence, both of these are dominated by the same kind of sources, i.e., blazars. 
Among the 133 sources in the \cite{Lister:2005p261}  MOJAVE sample,  94 (71\,\%) are in our sample as well.
Both samples, and the results derived from them, can hence be directly compared, as we do throughout this paper.

\section{Observations and Data Reduction}
\label{Obs}

The observations were performed with the XPOL polarimeter \citep{2008PASP..120..777T} on the IRAM 30\,m Telescope, by making use of the Observatory's orthogonal linearly polarized A100 and B100 heterodyne receivers tuned at 86\,GHz (3.5\,mm).
The main observing block was performed on July 2005.
To complete the list of observed sources, a small number of measurements had to be performed in a time block in September 2005 or at different epochs between 2006 and 2009 within different observing programs; see Table~\ref{T2}.

All observations were performed under the standard XPOL set-up and calibration scheme discussed in  \citet{2008PASP..120..777T}. 

Every XPOL polarization measurement consisted of a series of wobbler switching on-offs with typical integration times $>4$\,min., depending on the source's total flux density. 
Every polarization integration was preceded by a cross-scan pointing of the telescope and an amplitude, phase, and decorrelation loss polarization-specific calibration measurement \citep[see][]{2008PASP..120..777T}.

To estimate the remaining instrumental polarization to be subtracted from the data (\emph{a posteriori}), we made measurements of strong, compact and unpolarized sources at almost every observing epoch.
For 86\,GHz observations at the 30\,m Telescope, the HPBW of the beam is $\sim$28", 
hence an object with an apparent size of ($\aplt$4") can be safely considered as compact for our purposes. 
As a rule, we used Mars, Uranus, and Neptune as instrumental polarization calibrators.
When not available, other planets with up to $\sim$12" at the time of the observation (e.g., Venus), the compact H\,II regions W3\,OH, K3-50A ,and NGC\,7538, and the planetary nebula NGC\,7027, were observed for cross-check. 
 
The output of every polarization observation consists of a wide band (640\,MHz) spectral measurement from both the A100 and B100 receivers, together with the real and imaginary part of their cross-correlation.
These four observables are needed to recover the 4 Stokes parameters for each measurement as described in \citet{2008PASP..120..777T}.
After the observations, the polarization calibration measurements were used to calibrate the amplitudes of the A100 and B100 measurements, their relative phase, and the decorrelation losses.
Such calibrations were applied within the MIRA and CLASS software packages inside GILDAS.
The instrumental polarization for every observing epoch ($|Q_{\rm{i}}|\aplt$2\,\%, $|U_{\rm{i}}|\aplt$0.5\,\%, and $|V_{\rm{i}}|\aplt$0.5\,\%) was then estimated and removed from the data.
The $C_{\rm{Jy/K}}=6.4$\,Jy/K calibration factor for 86\,GHz observations at the 30\,m Telescope \citep[e.g.,][]{Agudo:2006p203} was also applied to the total flux density measurements ($S_{86}$).
At this stage, the uncertainties in every specific measurement only account for statistical uncertainties on the average of data along the spectral bandwidth.
The final errors in the total flux density measurements were computed by adding quadratically a 5\,\% systematic factor coming from the uncertainties in $C_{\rm{Jy/K}}$ \citep{Agudo:2006p203}.
In this stage, the remaining polarimetric non-systematic errors, which determine our polarimetric precision, were estimated from the dispersion in the  $Q$, $U$, and $V$ Stokes parameters from our main unpolarized calibrators, i.e. Mars and Uranus.
These dispersion estimates ($\Delta Q_{\rm{i}}=$0.5\,\%,  $\Delta U_{\rm{i}}=$0.3\,\%, and  $\Delta V_{\rm{i}}=$0.1\,\%) were also added in quadrature to provide the final uncertainty for every polarization measurement. 
The latter translate into final polarization-uncertainty medians of $\Delta \tilde{m}_{\rm{L}}\approx$0.53\,\%,  $\Delta \tilde{\chi}\approx$5\,$^{\circ}$, and  $\Delta \tilde{m}_{\rm{C}}\approx$0.2\,\% for the linear polarization degree  ($m_{\rm{L}}$), the linear polarization electric vector position angle ($\chi$), and the circular polarization degree ($m_{\rm{C}}$), respectively.
The uncertainties of our best measurements are dominated by $\Delta Q_{\rm{i}}$,  $\Delta U_{\rm{i}}$, and  $\Delta V_{\rm{i}}$, and hence their final uncertainties are $\Delta{{m}_{\rm{L}}}\approx$0.5\,\%,  $\Delta{\chi}\approx$1.5\,$^{\circ}$, and  $\Delta{m}_{\rm{C}}\approx$0.1\,\%.

\section{Results and Discussion}
\label{Res}

In Table~\ref{T2} we present the observing epoch and integration time for every source in our sample,  as well as the observational results, expressed in terms of $S_{86}$, $m_{\rm{L}}$, $\chi$, and $m_{\rm{C}}$.
A $3\sigma$ upper limit in both $m_{\rm{L}}$ and $m_{\rm{C}}$ is given whenever the measurement did not exceed such $3\sigma$ value.
Note that whereas linear polarization was detected from most sources in our sample (76\,\%), circular polarization could be detected only for a small fraction of them (6\,\%).

Statistical analysis and discussion of the relevant aspects regarding these data are presented in the following sub-sections.

\subsection{Total Flux Density}

\subsubsection{Total Luminosity}
\label{Lum}
\begin{figure}
   \centering
   \includegraphics[width=8.5cm]{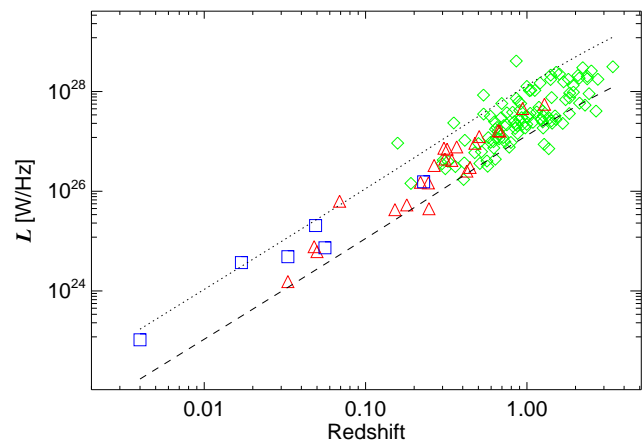}
   \caption{86\,GHz luminosity as a function of redshift. The dashed line indicates the luminosity for observer's frame flux density $S_{86}=0.5$\,Jy, whereas the dotted line is for $S_{86}=5$\,Jy. Diamonds symbolize quasars, triangles denote BL~Lacs, and squares {correspond to} radio galaxies.}
   \label{L_z_QBG}
\end{figure}

\begin{figure}
   \centering
   \includegraphics[width=8.5cm]{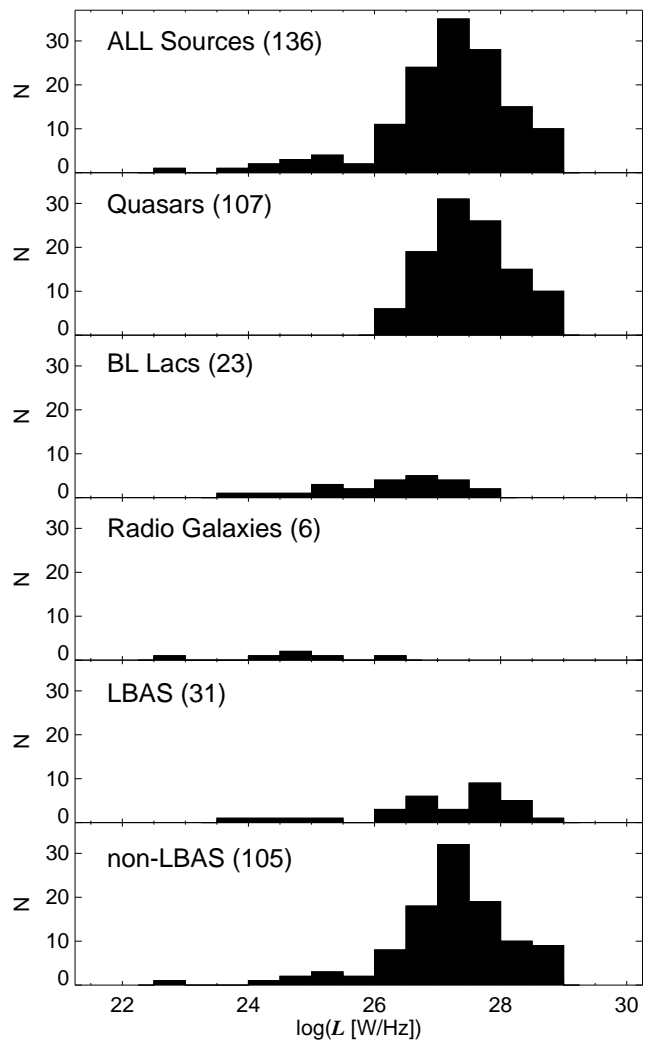}
   \caption{Distribution of  86\,GHz luminosity for all, quasars, BL~Lac sources, radio galaxies,  LBAS sources (i.e., those sources in our entire sample that were detected by {\emph Fermi}-LAT during its first 3 months of operation \citep[Tables~1 and 2 of][]{Abdo:2009p7779}), and non-LBAS sources. {Numbers in parentheses denote sample sizes of sources with known redshift.}}
   \label{LQBG}
\end{figure}

Fig.~\ref{L_z_QBG} shows the 86\,GHz luminosity ($L=4 \pi d_{L}^{2} S_{86}(1+z)^{-1}$, where $d_{L}$ is the luminosity distance for a $H_o=71$\,km\,s$^{-1}$\,Mpc$^{-1}$, $\Omega_{m}=0.27$ and $\Omega_{\Lambda}=0.73$ cosmology, used hereafter) as a function of redshift for our sample, with most --- all but 19 sources --- above the $S_{86}=0.5$\,Jy threshold in the observer's frame.
Only 6 sources in our sample displayed $S_{86}>5$\,Jy. 
As expected by the dependence of $L$ on cosmological distance, quasars, the most distant AGN class, show the largest luminosities (with median $\tilde{L}_{Q}=3.1\times10^{27}$\,W/Hz), followed by BL~Lac objects ($\tilde{L}_{B}=2.7\times10^{26}$\,W/Hz), and radio galaxies ($\tilde{L}_{G}=3.8\times10^{24}$\,W/Hz), the closest AGN (see also Fig.~\ref{LQBG}).

The LBAS sources in our sample show an $L$ distribution which is apparently different from that of non-LBAS sources (Fig.~\ref{LQBG}).
The median and peak luminosities of LBAS sources ($\tilde{L}_{LBAS}=2.4\times10^{27}$\,W/Hz, ${L}_{LBAS}^{peak}$ within $[3,10]\times10^{27}$\,W/Hz) are larger than those for non-LBAS ones ($\tilde{L}_{non-LBAS}=1.9\times10^{27}$\,W/Hz, ${L}_{non-LBAS}^{peak}$ within $[1,3]\times10^{27}$\,W/Hz).
However, the Kolmogorov--Smirnov (K-S) test does not give a sufficiently high confidence level (i.e., 70.7\,\%) to conclude that both distributions are selected from different parent distributions\footnote{As usual, only confidence levels $\ge95.0$\,\% will be considered sufficiently high to claim that two distributions are significantly different throughout this paper.}.

\begin{figure}
   \centering
   \includegraphics[width=8.5cm]{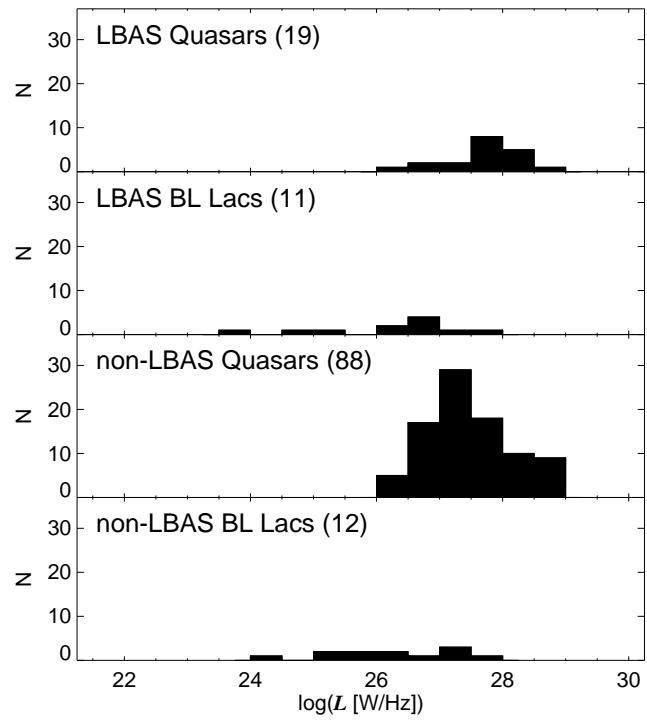}
   \caption{Distribution of  86\,GHz luminosity for LBAS quasars, LBAS BL~Lacs, non-LBAS quasars, and non-LBAS BL~Lacs in our source sample. {Numbers in parentheses denote sample sizes of sources with known redshift.}}
   \label{LQBLN}
\end{figure}

Comparison of the luminosity of $\gamma$-ray classes of quasars and BL~Lacs separately (Fig.~\ref{LQBLN}) shows that LBAS quasars in our sample are apparently more luminous ($\tilde{L}_{LBAS quasars}=7.0\times10^{27}$\,W/Hz, ${L}_{LBAS quasars}^{peak}$ within $[3,10]\times10^{27}$\,W/Hz) than non-LBAS quasars ($\tilde{L}_{non-LBAS quasars}=2.7\times10^{27}$\,W/Hz, ${L}_{non-LBAS quasars}^{peak}$ within $[1,3]\times10^{27}$\,W/Hz), whereas BL~Lacs show similar median luminosities and luminosity distributions independently of their $\gamma$-ray properties.
Indeed, the K-S test shows that the distributions of LBAS and non-LBAS quasars are significantly different (within a 99.0\,\% confidence level of both being drawn from different parent distributions), whereas LBAS and non-LBAS BL~Lacs are not (confidence level only 13.8\,\%).

Note that, for the computation of the confidence level of rejection of the null hypothesis in the K-S test, the effective number of data points ($N_{e}$)\footnote{$N_{e}=N_{1}N_{2}/(N_{1}+N_{2})$, where $N_{i}$ is the number of points on each sample, as defined by \citet{Press:1992}}, must be $\ge4$ to rely on the K-S test \citep{Press:1992}. 
All K-S tests presented in this paper, including the one for LBAS and non-LBAS BL~Lacs mentioned above (with $N_{e}=5.7$), fulfill this requirement.

\citet{Lister:2009p6166} found that among the 26 MOJAVE AGN contained in the LBAS sample, quasars have, on average, faster superluminal features than non-LBAS quasars. 
They also report evidence of faster superluminal proper motions in  $\gamma$-ray variable LBAS AGN with regard to non variable ones. 
\citet{Kovalev:2009p6218} report a correlation of the  $\gamma$-ray photon flux of 77 LBAS sources with contemporaneous VLBA or single dish flux density at 15\,GHz.
They also show that LBAS sources observed in their 15\,GHz VLBA sample display, in general, larger compact radio flux densities, larger brightness temperatures of their parsec-scale cores, and more active radio states than non-LBAS sources. 
These results are in good agreement with previous findings based on EGRET data suggesting that $\gamma$-ray bright blazars posses larger Doppler factors than weak ones \citep{2001ApJ...556..738J,Jorstad:2001p5655,Lahteenmaki:2003p5657,Kellermann:2004p3406}.
However, \citet{Lister:2009p6166} also suggest that the relativistic Doppler factor is not the sole parameter controlling the  $\gamma$-ray properties of blazars.

Our results are consistent with previous claims suggesting that $\gamma$-ray bright blazars have larger Doppler factors than weak $\gamma$-ray blazars, but we can only confirm the larger luminosity of $\gamma$-ray bright quasars against $\gamma$-weak quasars, and no significant difference for BL~Lacs.

We note, though, that our 86\,GHz observations were not performed contemporaneously with the {\emph Fermi}-LAT  $\gamma$-ray observations by \citet{Abdo:2009p7779}, and the luminosity distributions of LBAS and non--LBAS sources presented here (specially for BL~Lacs) may appear broader by 3\,mm source variability, thus hiding actual trends (see \S~\ref{Var}).

\subsubsection{15\,GHz to 86\,GHz Spectral Index}
\label{alpha}

The spectral index between two observing frequencies ($\alpha_{\nu_{1}{\rm,}\nu_{2}}=log(S_{\nu_{1}}/S_{\nu_{2}})/log({\nu_{1}}/{\nu_{2}})$) in the radio--mm spectral range provides information about the synchrotron opacity between such observing frequencies (${\nu_{1}{\rm,}\nu_{2}}$) affecting the radiation from an emitting jet region.
For the 15\,GHz and 86\,GHz emission in radio loud AGN, the emitting regions are not co-spatial in general (see Section~\ref{Intr}). 
Hence, for most sources, the $\alpha_{15{\rm,}86}$ determination is biased towards larger (more optically thick) spectral indices.
This is because the innermost 86\,GHz jet emitting regions are usually opaque to 15\,GHz radiation.

With this in mind, we performed an $\alpha_{15{\rm,}86}$ study of our main subsamples \emph {only} to study the relative differences among them.
For this, we used only those sources in our sample with available 15\,GHz total flux densities from integrated intensities of MOJAVE VLBA images\footnote{\tt http://www.physics.purdue.edu/MOJAVE}.
To try to avoid biasing the computation of $\alpha_{15{\rm,}86}$ owing to source variability, we selected, for each source, the MOJAVE observation closest in time to our 86\,GHz measurement.

\begin{figure}
   \centering
   \includegraphics[width=8.5cm]{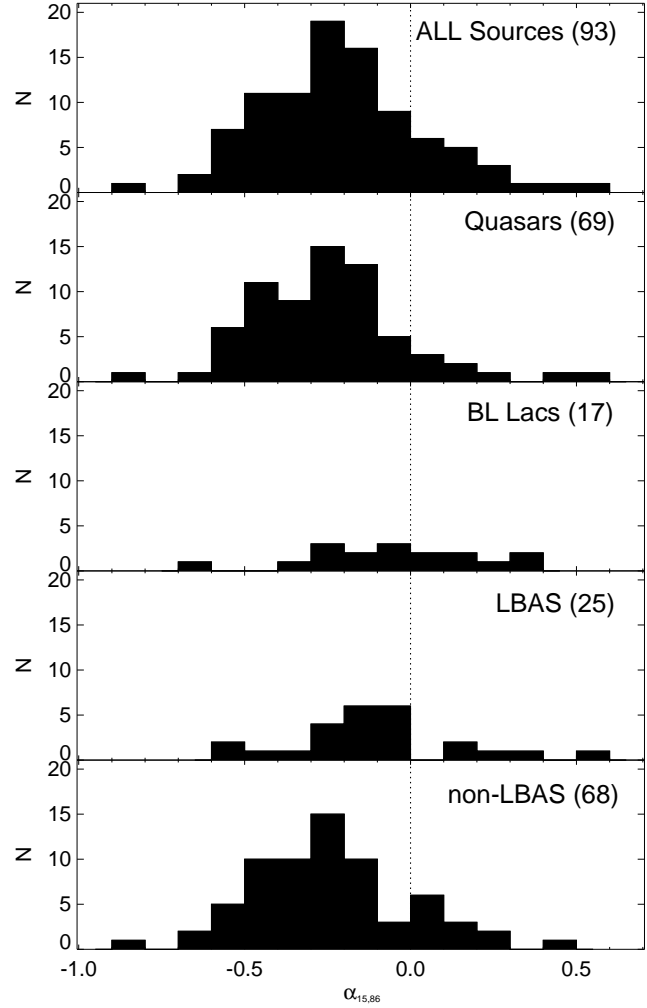}
   \caption{Distribution of 15\,GHz to 86\,GHz spectral {indices} ($\alpha_{15{\rm,}86}$) for all sources in both the MOJAVE and our sample, and their corresponding quasar, BL~Lac, LBAS, and non-LBAS sub-samples. The 15\,GHz total flux density was taken from integrated intensity of MOJAVE images. For each source, the closest 15\,GHz observation to our 86\,GHz measurement was selected. {Numbers in parentheses denote sample sizes.}}
   \label{SPIND}
\end{figure}

The $\alpha_{15{\rm,}86}$ distributions for each one of the source samples considered in this paper are shown in Fig.~\ref{SPIND}.
Even when affected by the above mentioned bias, the spectral indices for the whole source sample are distributed towards flat and optically thin spectral indices (with $\alpha_{15{\rm,}86}$ median $\tilde\alpha_{15{\rm,}86}=-0.23$, as expected from the definition of the sample and the high observing frequency), with a small fraction of sources (18\,\%) showing flat to optically thick spectral index.
Quasars and non-LBAS sources posses $\alpha_{15{\rm,}86}$ distributions similar to those of the entire source sample ($\tilde\alpha_{15{\rm,}86}^{\rm{quasar}}=-0.27$, $\tilde\alpha_{15{\rm,}86}^{\rm{non-LBAS}}=-0.26$).
However, both BL~Lacs and LBAS sources clearly distribute spectral indices more uniformly over an almost symmetric range of $\alpha_{15{\rm,}86}$ in $[-0.7,0.6]$.
This points out that BL~Lac and LBAS sources tend to show considerably flatter mm spectra (partially optically thick) than quasars and non-LBAS sources, respectively.
The significance of this result is guaranteed by the K-S test, which points out that the $\alpha_{15{\rm,}86}$ distributions for quasars and BL~Lacs in Fig.~\ref{SPIND} are different (at 99.4\,\% confidence level), as well as those for LBAS and non-LBAS subsamples (within a 96.9\,\% confidence level).
We note that quasars in the LBAS subsample are dominated by sources with spectral indices above $\tilde\alpha_{15{\rm,}86}^{\rm{quasar}}$, whereas LBAS BL~Lacs have a more spread $\alpha_{15{\rm,}86}$ distribution along the ranges of the entire BL~Lac subsample.

In principle, the spectral differences shown by radio loud quasars and BL~Lacs in our analysis would be expected if jets in BL~Lacs would be preferentially oriented towards smaller viewing angles (with regard to the line of sight).
In this case, the integrated radiation coming from different spectral components is expected to flatten the spectrum \citep[e.g.,][]{1980ApJ...238L.123C}.
However, recent work \citep{Hovatta:2009p3860,Pushkarev:2009p9412} has concluded that flat spectrum radio quasars have significantly smaller viewing angles than BL~Lac objects, which is expected to produce the reverse spectral behavior from that observed by us.
In this case, we can attribute more confidently the average spectral differences shown by quasars and BL~Lacs in Fig.~\ref{SPIND} to the larger cosmological redshifts of quasars (see Fig.~\ref{LQBG}), which shift their spectra to lower frequencies in the observer's frame.
This preferentially reveals the optically thin part of the synchrotron spectrum of quasars with regard to BL~Lacs.
Also, we cannot rule out differences in the intrinsic spectral properties of the emitting particle populations between these two source classes.

The means of the viewing angles of LBAS and non-LBAS sources contained in Table~1 of \citet{Pushkarev:2009p9412} are not statistically significant according to their Student's T-test.
This does not allow us to attribute the spectral differences between these two subsamples to preferential alignment of $\gamma$-ray bright blazar jets with the line of sight, although this possibility cannot be ruled out based on only this information.
Intrinsic differences in the properties of the synchrotron-emitting particle populations of these two subsamples, as well as source redshifts, can, in principle, be relevant as well.

\subsection{Linear Polarization}

Fractional linear polarization at 86\,GHz ($m_{\rm{L}}$) was detected for 110 sources, 76\,\% of our sample.
When comparing the median values of $m_{\rm{L}}$ ($\tilde{m}_{\rm{L}}$) for the major optical classes, we find that BL~Lac objects, with $\tilde{m}_{\rm{L}}=4.4$\,\%, are more strongly polarized than quasars, with $\tilde{m}_{\rm{L}}=3.1$\,\%. 
To avoid biassing the $\tilde{m}_{\rm{L}}$ values, $3\sigma_{m_{\rm{L}}}$ upper limits were considered for the cases when $m_{\rm{L}}$ could not be detected. 
Otherwise, if the non-detections are not accounted for when calculating the medians, $\tilde{m}_{\rm{L}}$ would be overestimated by $\sim1$--$1.5$\,\%\footnote{Note that assigning different values to the non-detected $m_{\rm{L}}$ does not have an effect on the resulting median, provided that all estimates of $m_{\rm{L}}$ are smaller than $\tilde{m}_{\rm{L}}$, as is the case for all considered subsamples.}.
The difference between the quasar and the BL~Lac $m_{\rm{L}}$ distributions is confirmed by the Gehan's generalized Wilcoxon (GGW) test at a 95.2\,\% confidence level.
To perform this test, which takes into account both detections and upper limits, we used the ASURV~1.2 survival analysis package \citep[see ][ and references therein]{Lavalley:1992}.

This result (i.e., quasars less polarized at 86\,GHz, in general, than BL~Lacs) does not seem consistent with quasars having a significantly thinner synchrotron spectrum than BL~Lacs between 15 and 86\,GHz. 
This is expected to reduce $m_{\rm{L}}$ for BL~Lacs with regard to quasars owing to stronger synchrotron self-absorption in BL~Lacs.
Hence, the radio-to-mm spectral properties of quasars and BL~Lacs cannot explain their overall linear polarization properties.  
In contrast, an explanation comes from recent evidence that the viewing angle of jets in quasars is smaller than that in BL~Lacs \citep{Hovatta:2009p3860,Pushkarev:2009p9412}.
If either the magnetic field is not homogeneously distributed along the jet or the jet has prominent non-axysimmetry, lower polarization degree is expected from sources better oriented to the line of sight  (i.e., quasars) owing to cancellation of orthogonal polarization components.

\begin{figure}
   \centering
   \includegraphics[width=8.5cm]{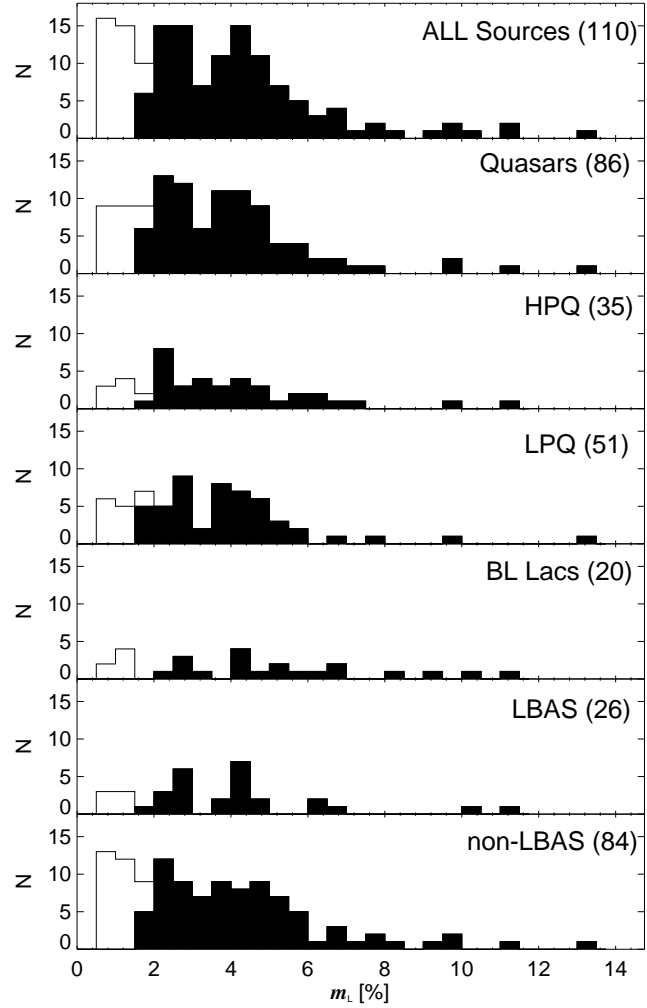}
   \caption{Distribution of 86\,GHz fractional linear polarization for: all sources in the entire sample, quasars, high polarization quasars (HPQ), low polarization quasars (LPQ), BL~Lac objects, radio galaxies, LBAS, and non-LBAS sources (from top to bottom). N is the number of sources in 1\,\% wide bins. {Unfilled} areas correspond to non-detections.
Note that since we have chosen to consider $3\sigma$ upper limits with $\sigma_{m_{\rm{L}}}\approx0.5$\,\%, no data with $m_{\rm{L}}\aplt1.6$\,\% {are} available. {Numbers in parentheses denote sizes of detected $m_{\rm{L}}$ samples.}}
   \label{mALL}
\end{figure}

Interestingly, the $m_{\rm{L}}$ distribution of the entire source sample is double peaked (Fig.~\ref{mALL}), with the first peak at  $m_{\rm{L}}\approx2.5$\,\%, and the second one at $m_{\rm{L}}\approx4$\,\%.
At lower polarization degrees, a similar bimodal distribution was previously observed by \cite{Lister:2005p261} for the cores of quasars and the integrated polarization degree of EGRET-detected blazars.
Fig.~\ref{mALL} shows that this double peak in our data might come from the quasar subsample, which dominates the overall sample.

To test whether this dichotomy is produced by the presence of radio loud quasars with high optical polarization (HPQ, $\ge3$\,\%) and radio quasars with low optical polarization (LPQ, $<3$\,\%) in our sample, we have also analyzed these two subsamples separately (see Fig.~\ref{mALL}).
We considered HPQ as those sources in the \citet{VeronCetty:2006p4900} catalog with such a label, and LPQ as the remainder of sources in that catalog.
Simple inspection of Fig.~\ref{mALL} apparently shows that if there is a physical meaning on the two--peaked quasar distribution, it cannot be because of HPQ producing the higher $m_{\rm{L}}$ peak and the LPQ the one at lower $m_{\rm{L}}$.
Indeed, there is no statistically significant difference between the HPQ and LPQ $m_{\rm{L}}$ distributions that could confirm such a hypothesis (the GGW test gives only a 21.4\,\% confidence).

A similar double peak is seen in the $m_{\rm{L}}$ distribution for the LBAS source sample (Fig.~\ref{mALL}), whereas the second peak is not so evident in the non-LBAS source $m_{\rm{L}}$ distribution.
This might be interpreted as suggesting that this bimodal distribution comes from  $\gamma$-ray bright quasars. 
However, according to our GGW test, there is no significant difference between the $m_{\rm{L}}$ distributions of LBAS and non-LBAS sources (36.9\,\% confidence only).
 
Thus our analysis remains inconclusive about the origin of the apparent polarization degree dichotomy.

\subsubsection{86\,GHz to 15\,GHz Fractional Linear Polarization Ratio}
\begin{figure}
   \centering
   \includegraphics[width=8.5cm,clip]{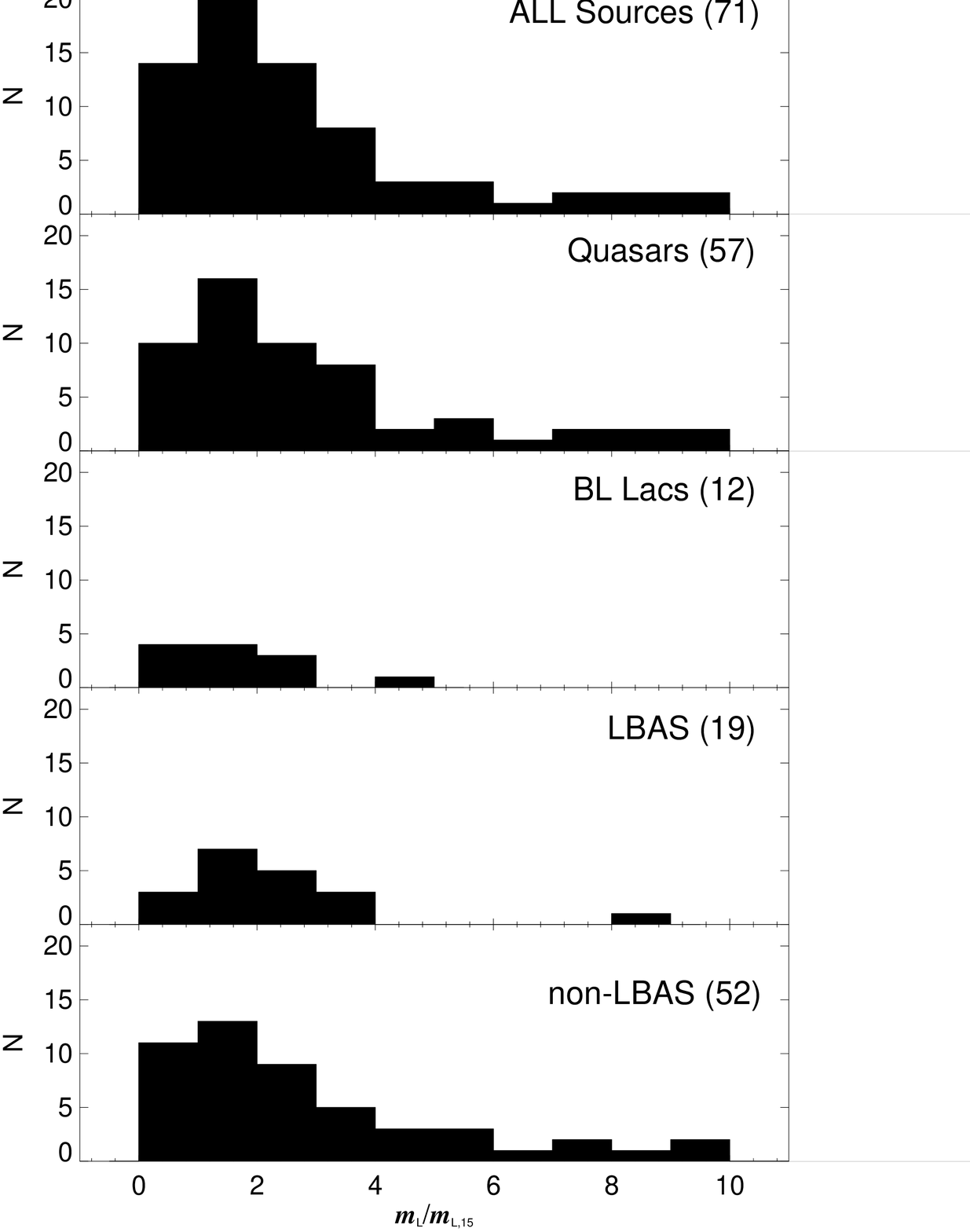}
   \caption{Distribution of 86\,GHz to 15\,GHz fractional linear-polarization ratio for sources with detected linear polarization both in our survey and in MOJAVE. Two sources with $m_{\rm{L}}$/$m_{\rm{L,15}}>17$ are not presented. The 15\,GHz linear polarization fraction was computed from measurements of integrated total flux density and {linearly polarized} flux density from 15\,GHz VLBI images listed in \cite{Lister:2005p261}. {Numbers in parentheses denote sample sizes.}}
   \label{ml_ml15_AQBLN}
\end{figure}

\begin{figure}
   \centering
   \includegraphics[width=8.5cm,clip]{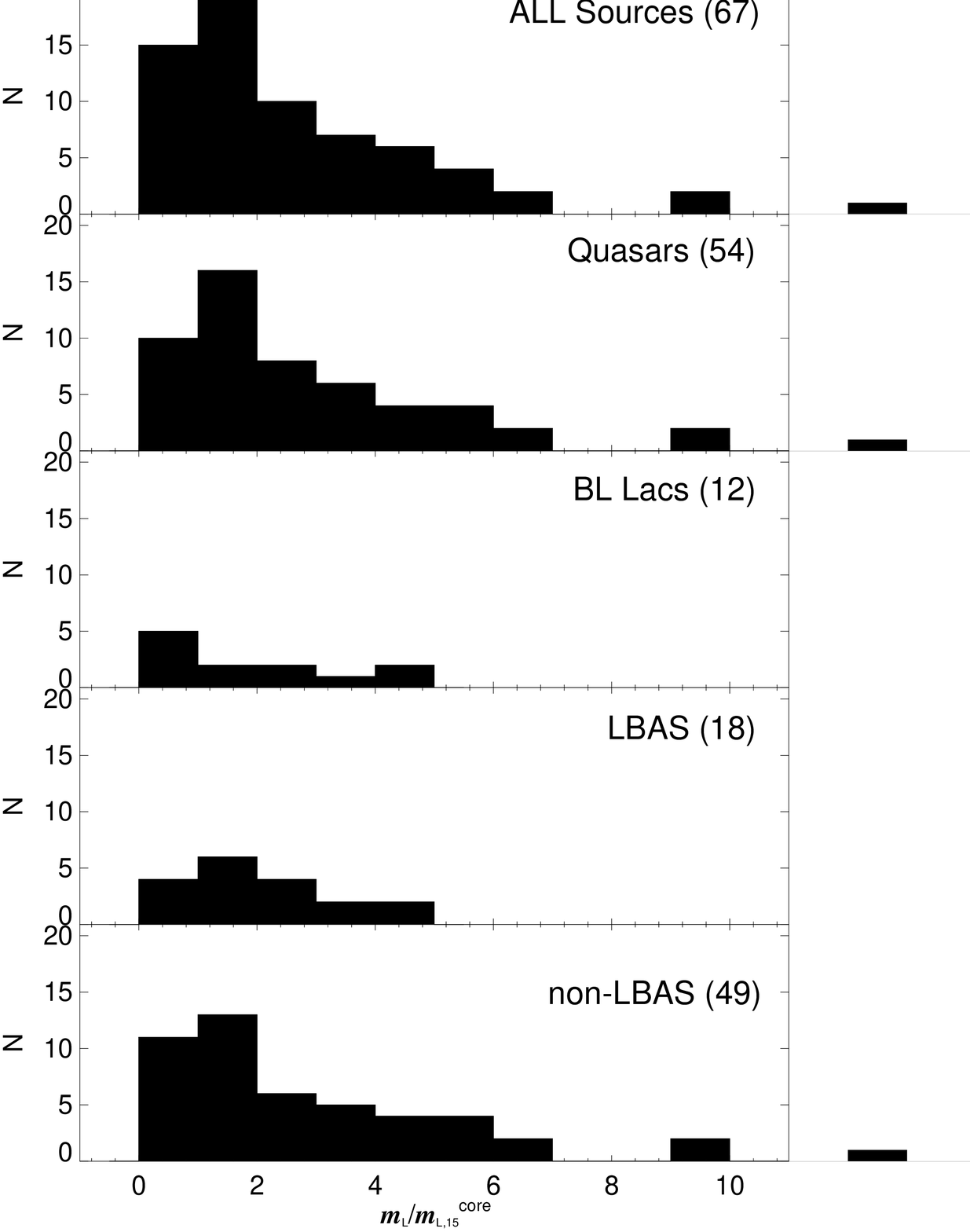}
   \caption{{Distribution of 86\,GHz to 15\,GHz fractional linear-polarization ratio for sources with detected linear polarization in our survey and those with detected core polarization in MOJAVE (i.e., $m_{\rm{L}}$/$m^{\rm{core}}_{\rm{L,15}}$). Two sources with $m_{\rm{L}}$/$m^{\rm{core}}_{\rm{L,15}}>14$ are not presented. The used $m^{\rm{core}}_{\rm{L,15}}$ measurements are those given by \cite{Lister:2005p261}. Numbers in parentheses denote sample sizes.}}
   \label{ml_ml15cor_AQBLN}
\end{figure}

The fraction of linear polarization detections in our sample is $\sim76$\,\%.
Considering that we adopted a $3\sigma$ criterion for detection (with median $3\tilde{\sigma}_{m_{\rm{L}}}\sim1.6$\,\% over the whole sample), this means that, on average, $\sim$76\,\% of our sources display $m_{\rm{L}}\apgt1.6$\,\% at 86\,GHz.
However, only 60\,\% of sources both in the MOJAVE and in our sample show 15\,GHz fractional linear polarization $\apgt1.6$\,\%.
This points to a general trend of blazars to increase their linear polarization degree with frequency, as previously suggested by \cite{Agudo:2006p203} and \cite{2007AJ.134.799J}.
Indeed, those sources for which linear polarization was detected both in our survey and in the integrated emission of the MOJAVE images \citep{Lister:2005p261} follow a clear general trend to show significantly larger fractional linear polarization at 86\,GHz than at 15\,GHz by a median factor $\approx2$ (Fig.~\ref{ml_ml15_AQBLN}).
The same factor $\approx2$ is obtained if our $m_{\rm{L}}$ data is compared with the linear polarization degree of the core at 15\,GHz ($m^{\rm{core}}_{\rm{L,15}}$) as measured by \citet{Lister:2005p261} (Fig.~\ref{ml_ml15cor_AQBLN}).

Note also that there is a prominent 22\,\% (24\,\%)  fraction of the entire source sample with larger $m_{\rm{L}}/m_{\rm{L,15}}$ ($m_{\rm{L}}/m^{\rm{core}}_{\rm{L,15}}$) ratio than $4$.
This tail of large 86\,GHz linear polarization excess seems to be present in all subsamples of optical and  $\gamma$-ray classes, although BL~Lacs do not show it clearly, perhaps because of the lower number of sources in that subsample.
 
This result may be explained by a combination of two phenomena, that the 86\,GHz emission in blazars comes from a region with greater degree of order of the magnetic field than the one at 15\,GHz, and that the 15\,GHz emission from blazars is affected by considerably greater Faraday depolarization relative to the 86\,GHz emission.

If the former hypothesis is true, it would imply that the magnetic field order in the inner regions of the relativistic jets is larger than in outer regions.
For this statement, we assume that the bulk of the 86\,GHz emission comes on average from inner jet regions where 15\,GHz emission is strongly affected by synchrotron opacity, hence locating the 15\,GHz core farther downstream \citep[e.g.,][]{2007AJ.134.799J}.

Despite the apparent differences between the source subsamples presented in each of Fig.~\ref{ml_ml15_AQBLN} and \ref{ml_ml15cor_AQBLN} (see also above), our K-S test does not allow us to claim any significant differences between the $m_{\rm{L}}/m_{\rm{L,15}}$ or $m_{\rm{L}}/m^{\rm{core}}_{\rm{L,15}}$ distributions either for quasars versus BL~Lacs (confidence level only 84.5\,\%, and 53.9\,\%, respectively), or for LBAS against non-LBAS sources (confidence level only 85.1\,\%, and 59.5\,\%, respectively).

\subsubsection{Total Luminosity vs. Linear Polarization}

\begin{figure}
   \centering
   \includegraphics[width=8.5cm]{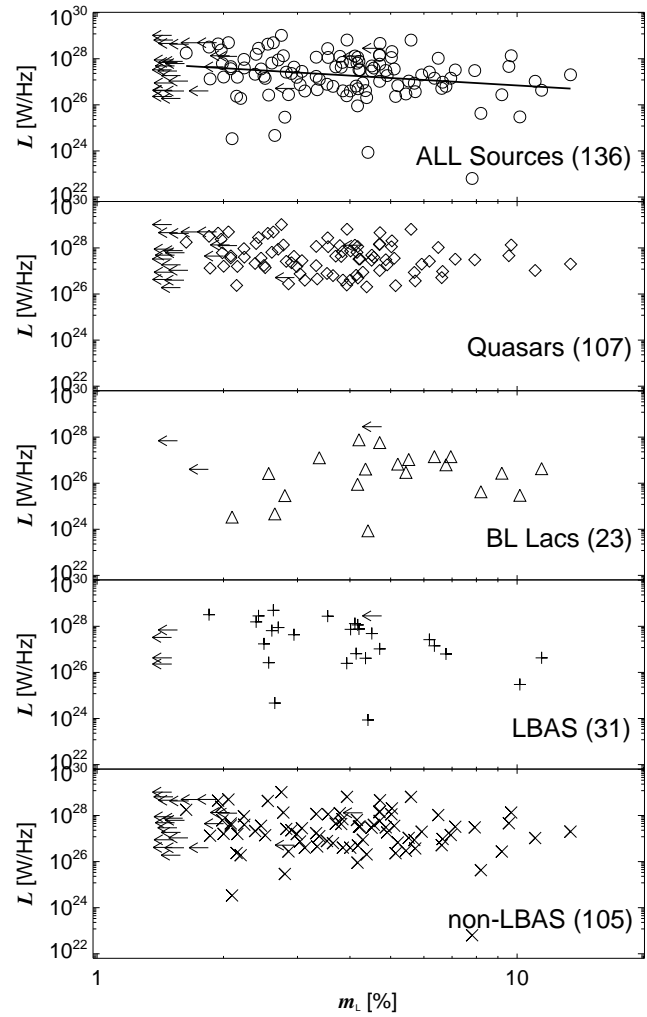}
   \caption{86\,GHz fractional linear polarization for sources {with known redshift} {for} the {entire} source sample, the quasar, the BL~Lac, LBAS, and {the} non-LBAS sub-samples (from top to bottom). {Arrows symbolize $m_{{\rm{L}}}$ upper limits. The continuous line in the upper plot symbolizes the result of a linear regresion. Numbers in parentheses denote sample sizes.}}
   \label{L_ml_AQBLN}
\end{figure}

Fig.~\ref{L_ml_AQBLN} shows the 86\,GHz luminosity {\emph vs.} 86\,GHz fractional linear polarization for sources with known redshift in the entire source sample and in the four major optical and $\gamma$--ray classes considered here.
Our correlation analysis for data containing upper limits --- performed with the ASURV~1.2 package by \citet{Lavalley:1992} --- shows significant correlation between $L$ and $m_{{\rm{L}}}$ for the whole source sample. 
This is supported by the results of the Cox, Kendall's $\tau$, and Spearman's $\rho$ tests. 
These indicate that $L$ and $m_{{\rm{L}}}$ are correlated at 98.8\,\%, 99.2\,\%, and 98.7\,\% confidence level, respectively.
Significant correlation is also found for the case of the non-LBAS subsample --- at 98.4\,\%, 97.0\,\%, and 96.7\,\% confidence levels, respectively.

However, for the quasar subsample --- with confidence levels of 87.7\,\%, 93.3\,\%, and 91.1\,\% for the Cox, the Kendall's $\tau$, and the Spearman's $\rho$ tests, respectively ---, no formally significant correlation is found.
Perhaps because of their decreased sizes, the correlation tests over the BL~Lac and the LBAS yield confidence levels of correlation of these subsamples that are even much lower.
 
Under the standard assumption that, in the absence of strong Faraday depolarization (which we assume to be the case for our 3.5\,mm observations, see Section~\ref{Intr}), higher degree of linear polarization reflects greater magnetic field order, we have found that the magnetic field order ---in our entire source sample--- increases with decreasing mm luminosity.

A reasonable hypothesis to explain this phenomenon would simply involve orientation and relativistic effects.
In principle, those sources whose jets are better oriented to the line of sight are expected to display larger luminosities (because of their larger Doppler factors) and also lower linear polarization degrees (because of cancellation of orthogonal polarization components along the line of sight).
To test this hypothesis, we used the ``variability Doppler factors'' given by  \cite{Hovatta:2009p3860} to compute the beaming corrected luminosities of the 70 sources in their and our sample.
The correlation analysis of all source types in Fig.~\ref{L_ml_AQBLN} points out that, although the reduced number of sources in every subsample tends to reduce the correlation between $L_{\rm{unbeamed}}$ and $m_{{\rm{L}}}$, the decreasing trend of $L_{\rm{unbeamed}}$ with increasing $m_{{\rm{L}}}$ still seems to be present.
Despite the reduction of data points, such anti--correlation is now statistically significant for the LBAS subsample (with 99.1\,\%, 99.5\,\%, and 98.4\,\% confidence level, from the Cox, Kendall's $\tau$, and Spearman's $\rho$ test, respectively), which do not allow us to support the beaming scenario as a reliable explanation for the $L$ and $m_{{\rm{L}}}$ anti--correlation.

This may have a plausible  alternative interpretation that, although still qualitative and speculative, may be a valid proposal to investigate further.
The AGN jet regions responsible for the 3\,mm emission are thought to be dominated by plasma dynamics, in contrast to the innermost Poynting flux-dominated jet region, where the collimation and acceleration is assumed to take place \citep[e.g.,][]{2007AJ.134.799J,2008Natur.452..966M,Marscher:2010p11374}.
It is known that increasing jet kinetic energies with regard to their surrounding low speed wind favor the generation of fluid instabilities, leading to turbulence, in the region separating the inner-fast jet and the outer-slow wind \citep[e.g.,][for different numerical setups for this scenario]{Mizuno:2007p242,Meliani:2009p8926}.
This picture fits into our observing results.
A faster jet would be more luminous, but would also produce more turbulence in the shear layer where the external wind penetrates the jet, hence reducing the magnetic field order and the linear polarization degree. 

\subsubsection{Linear Polarization Angle versus Jet Position Angle}
\label{Misal}

\begin{figure}
   \centering
   \includegraphics[width=8.5cm]{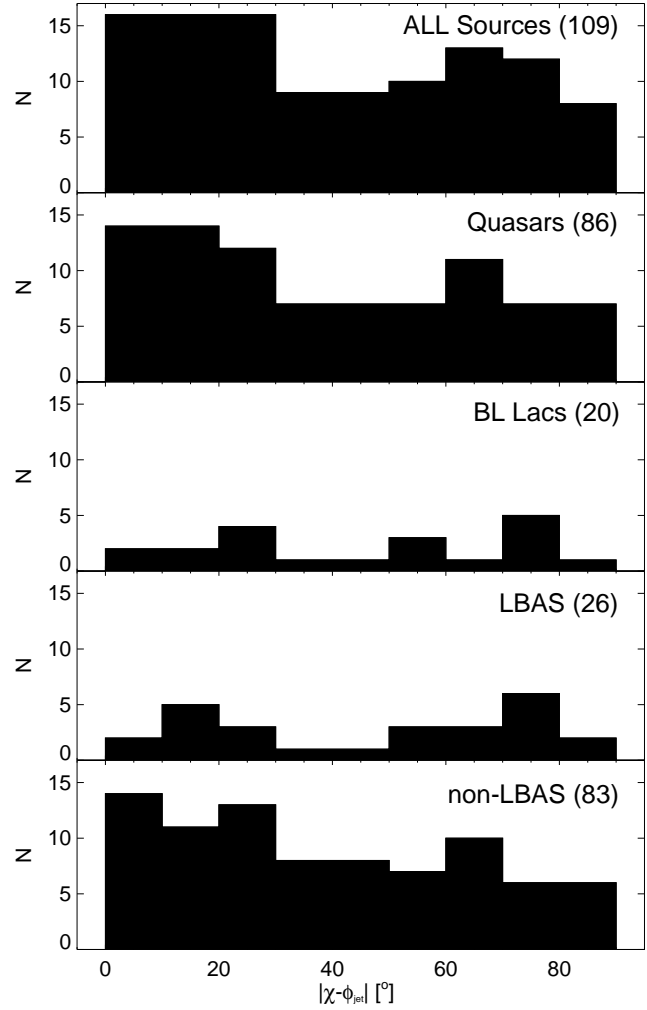}
   \caption{Distribution of misalignment between the 86\,GHz linear-polarization electric-vector position-angle ($\chi$, given in Table~\ref{T2}) and the jet structural position angle ($\phi_{\rm{jet}}$, given in Table~\ref{T1}). We present, from top to bottom, the {entire} source sample and the subsamples of quasars, BL~Lacs, LBAS, and non-LBAS sources.}
   \label{misal}
\end{figure}

In Fig.~\ref{misal} we show the distribution of misalignment of linear-polarization electric-vector position-angle ($\chi$, given in Table~\ref{T2}) with the jet structural position angle ($\phi_{\rm{jet}}$, given in Table~\ref{T1}), i.e., $|\chi-\phi_{\rm{jet}}|$, for the entire source sample, and the subsamples of quasars, BL~Lacs, LBAS, and non-LBAS sources.

Different angular resolutions achieved at different VLBI observing frequencies probe different jet regions that may show, in some particular cases, large $\phi_{\rm{jet}}$ differences \citep[see e.g.,][for the case of {NRAO~150} with $\Delta\phi_{\rm{jet}}\approx100^{\circ}$]{Agudo:2007p132}.
Opacity effects of the inner jet regions at low frequencies can also impede a reliable determination of $|\chi-\phi_{\rm{jet}}|$ if  $\phi_{\rm{jet}}$ is measured at $\nu<<86$\,GHz.
To try to avoid these biases as much as possible, we first searched for $\phi_{\rm{jet}}$ values in the 86\,GHz VLBI Survey by \citet{Lee:2008p301}.
When 86\,GHz VLBI images were not available, the 15\,GHz data from either the MOJAVE survey \citep[][preferentially]{Lister:2005p261} or those from the 2\,cm VLBA Survey \citep{Kellermann:2004p3406} were used.
Otherwise, a deeper search was done for every source from references 1 to 10 on Table~\ref{T1}. 
The higher frequency result thus found was then adopted.

Fig.~\ref{misal} shows a weak trend in the entire source sample of alignment of $\chi$ close to $0^{\circ}$--$30^{\circ}$ to $\phi_{\rm{jet}}$, such that $\chi$ lies almost parallel to the jet axis.
A similar pattern is found in the quasar and non-LBAS subsamples. 
In contrast, the values of $|\chi-\phi_{\rm{jet}}|$ in the BL~Lac and LBAS subsamples tend to lie within the ranges $10^{\circ}$--$30^{\circ}$ or $70^{\circ}$--$80^{\circ}$.
However, our K-S tests indicate that there is no significant difference between the quasar and the BL~Lac distributions (only 18.7\,\% confidence to come from the same parent distribution), and between the LBAS and non-LBAS distributions (only 63.9\,\% confidence).

\begin{figure}
   \centering
   \includegraphics[width=8.5cm]{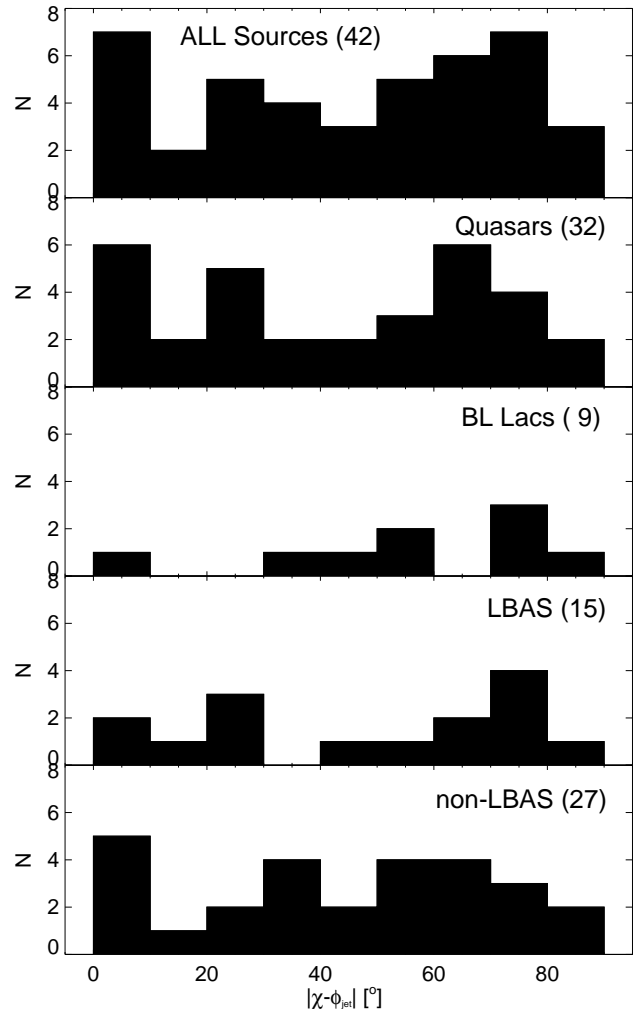}
   \caption{Same as Fig.~\ref{misal} but for sources with $S_{86}>1.5$\,GHz only.}
   \label{misal-1.5Jy}
\end{figure}

We therefore find no clear trend in any of the source samples considered in this work for $\chi$ to be aligned either parallel or perpendicular to $\phi_{\rm{jet}}$.
Actually, even for the entire source sample, which shows an apparent preference of aligned $\chi$ parallel to $\phi_{\rm{jet}}$, there is only a small excess of sources ($\sim$25\,\%) distributed at $|\chi-\phi_{\rm{jet}}|<30^{\circ}$.
Similar results are obtained from the quasar and LBAS distributions.
Even this $\sim25$\,\% excess must be interpreted with care.
When only highly polarized source states ($m_{\rm{L}}>3$\,\%) are considered -- which seems to reveal more clearly the $\chi$ to $\phi_{\rm{jet}}$ relation in the case of the quasar 3C~454.3 \citep{Jorstad:subm}  --, such excess is decreased to $\sim18$\,\%.
Moreover, Fig.~\ref{misal-1.5Jy} shows that, if the $|\chi-\phi_{\rm{jet}}|$ distributions are generated from a smaller number of sources (only those brighter than 1.5\,Jy at 86\,GHz), such an excess disappears completely. 
All this shows that if there is a preference to align $\chi$ parallel to the jets of the sources shown in Fig.~\ref{misal}, such a preference is intrinsically weak at 86\,GHz and it is also partially hidden by random changes in the $|\chi-\phi_{\rm{jet}}|$ distributions.

Theoretically, the electric vector position angle ($\chi$) of the linearly polarized emission from relativistic axisymmetric jets should be observed either parallel or perpendicular to the jet axis \citep[e.g,][]{Lyutikov:2005p321,Cawthorne:2006p409}.
However, the results from several observational attempts to confirm this bi-modality do not show a robust agreement among each other.
\citet{2000MNRAS.319.1109G}, through 5\,GHz VLBI observations, reported a tendency of the core and jets in a set of 25 BL~Lac objects to show $\chi$ either parallel (preferentially) or perpendicular to the jet direction.
However, on the one hand, \citet{2003ApJ...589..733P} found a strong tendency for the cores of their observed quasars to possess values of $\chi$ that are perpendicular to the jet axis, whereas no correlation was found for the jets in quasars, nor for the jets or cores in BL~Lac objects.  
On the other hand, \citet{Lister:2005p261} found the cores of sources in the MOJAVE sample to show $\chi$ values aligned preferentially parallel to the jet axis, which was found to be a particularly strong tendency for the case of BL~Lac objects. 
Such a preference was not found in the jets of their sample, although BL~Lacs did show a better tendency to have $\chi$ aligned with the jets.
The apparent lack of consensus from these studies might be the result of the well known frequency dependent Faraday rotation of $\chi$ \citep[e.g.,][]{Zavala:2004p138}.

Our 86\,GHz polarimetric observations show that there is not a clear trend in our data for $|\chi-\phi_{\rm{jet}}|$ to be distributed either near $\sim0^{\circ}$ or near $\sim90^{\circ}$.
This difference between our results and those by \citet{Lister:2005p261} cannot be attributed to the use of $\phi_{\rm{jet}}$ measured at 86\,GHz in our analysis.
Indeed, for the 40 sources in our sample having measurements of both $\phi_{\rm{jet,}>{\rm{43\,GHz}}}$ and $\phi_{\rm{jet,15\,GHz}}$, both magnitudes are in general good agreement (with Spearman's correlation rank $\rho=0.91$ at 99.9\,\% confidence, and mean of $\phi_{\rm{jet,}>{\rm{43\,GHz}}} - \phi_{\rm{jet,15\,GHz}}\approx9^{\circ}$).  
Also, when  $\phi_{\rm{jet,15\,GHz}}$ is used instead of $\phi_{\rm{jet,}>{\rm{43\,GHz}}}$ for these 40 sources, the distribution in Fig.~\ref{misal} does not show noticeable changes.

The median (maximum) uncertainty in the determination of $\chi$ for those sources in our sample detected in linear polarization is $\bar{\delta\chi}\approx4^{\circ}$ ($\delta\chi^{\rm{max}}\approx9^{\circ}$). 
Thus, these uncertainties cannot explain the lack of a strong trend in our $|\chi-\phi_{\rm{jet}}|$ distributions for $\chi$ to be aligned either parallel or perpendicular to $\phi_{\rm{jet}}$.

The large amplitude and rapid linear polarization variability at mm \citep[e.g.,][]{2007AJ.134.799J} and radio wavelengths \cite[e.g.,][and references therein]{2003ApJ...586...33A}, for which blazars are well known, is certainly an unavoidable effect both in our $|\chi-\phi_{\rm{jet}}|$ study and in those by previous authors (see \S~\ref{Var}). 
However, we do not find strong arguments supporting a scenario that hides a clear $|\chi-\phi_{\rm{jet}}|$ trend in our data when allowing for it in previous radio surveys. 
Considerably larger $\chi$ variability amplitude, or a larger probability to find sources in high variability states at mm wavelengths could certainly explain this scenario, but there is still no strong observational support for it, as far as the authors know.

There is though a likely explanation for the differences between our results and those by \citet{Lister:2005p261} and \citet{2003ApJ...589..733P}. 
Our data, which are essentially free of Faraday rotation (see Section~\ref{Intr}), suffer from prominent synchrotron opacity well upstream the 15\,GHz and 5\,GHz cores, hence reflecting the properties of inner jet regions than the sections of the jet that radiate predominantly at radio frequencies.
Apart from other biases, differences on the $|\chi-\phi_{\rm{jet}}|$ distributions a mm and radio wavelengths could be reflecting the properties of the regions from where most of the emission at such wavelengths comes from.

There is an increasing number of AGN jets where large mm-$\chi$ rotations ($>>90^{\circ}$) with typical time scales of days to months are detected either in the cores or in bright moving features \citep[e.g.,][]{2007ApJ...659L.107D, Larionov:2008p338,Jorstad:subm}.
Successful explanation of these phenomena requires non-axisimmetric jet dynamics such as emitting features that propagate down the jet along helical paths \citep{2008Natur.452..966M, Marscher:2010p11374}.
Gradients of axisymmetric Faraday rotation screens along the path of moving jet features may also produce large $\chi$ rotations at 7\,mm \citep[e.g.,][]{2000Sci...289.2317G,Gomez:2001p201}, but much larger than typical RM values ($>>10^4$\,rad\,m$^{-2}$) are required to produce rotations at 3\,mm to explain Figs.~\ref{misal} and \ref{misal-1.5Jy}.

\subsubsection{AGN as Source of Contamination of the 86\,GHz Linear Polarization CMB}
\label{CMB}

\begin{figure}
   \centering
   \includegraphics[width=8.5cm,clip]{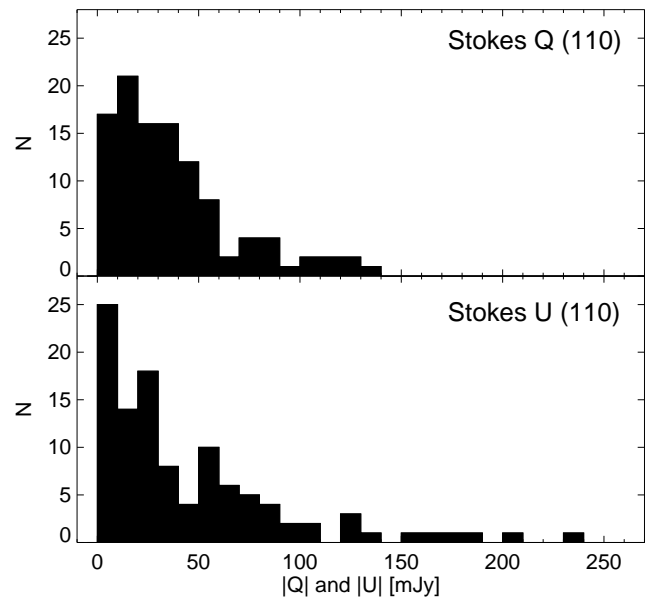}
   \caption{Distribution of $|Q|$ (top) and $|U|$ (bottom) Stokes parameters for {sources with detected} linear polarization. Four sources with $|Q|$ and/or $|U|$ in the range $[300,1000]$\,mJy are not shown.}
   \label{QU_ALL}
\end{figure}

Fig.~\ref{mALL} show that linear polarization of flat spectrum radio loud AGN is larger than previously expected from centimeter wavelength surveys \cite[e.g.,][]{2003ApJ...589..733P,Ricci:2004p161,Lister:2005p261}.
This is clearly indicated by the excess by a factor of $\sim2$ of 86\,GHz linear polarization relative to 15 GHz linear polarization (Fig.~\ref{ml_ml15_AQBLN}), which points out that simple extrapolation from measurements at cm wavelengths systematically underestimates the CMB-foreground polarization contribution of AGN at 3\,mm wavelength.
The median fractional linear polarization of detected sources is $\sim4$\,\% with a considerable population ($27$\,\%) showing linear polarization $>5$\,\% (Fig.~\ref{mALL}).
Fig.~\ref{QU_ALL}  shows that, at 3\,mm wavelengths, linear polarization in radio loud AGN may reach significant values relative to the CMB contribution for high sensitivity measurements. 
Indeed, 62\,\% (20\,\%) of our linear polarization detected sources show $|Q|$ or $|U|>50$\,mJy ($>100$\,mJy).

Hence, our observing results in Table~\ref{T2} are relevant, and can be used, to estimate the 3\,mm AGN contribution to linear polarization CMB measurements, such as those performed by the Planck satellite\footnote{\tt http://www.rssd.esa.int/index.php?project=Planck}. 
A new epoch of observations, over an improved flux density limited, complete sample, is already planned for mid 2010. This will provide contemporaneous measurements with those of Planck, as well as the possibility to study the effects of linear polarization variability on the CMB radio loud AGN foregrounds.

\subsection{Circular Polarization}
\label{mC}

\begin{figure}
   \centering
   \includegraphics[width=8.5cm]{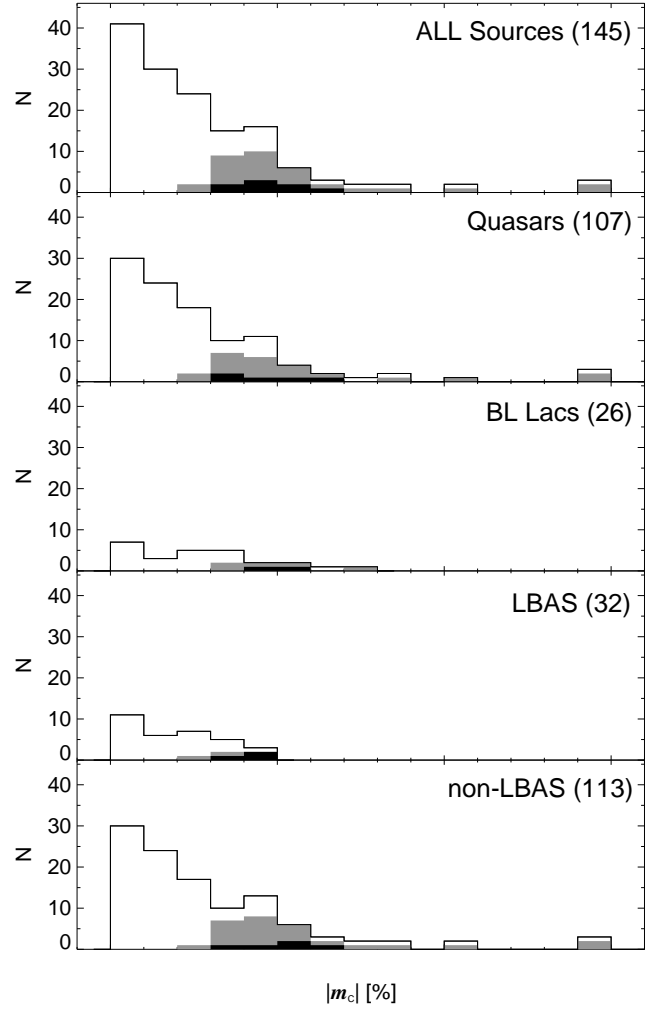}
   \caption{Distribution of the absolute value of circular polarization for all major samples (labeled for each sub-plot) considered in this study. Black areas correspond to $m_{\rm{C}}$ detections at $\ge3\sigma$. Grey shaded areas indicate observing results with $\ge2\sigma$, whereas unshaded areas symbolize all $m_{\rm{C}}$ measurements, independently of their significance.}
   \label{mcALL}
\end{figure}

The distribution of $|m_{\rm{C}}|$ for all major samples considered in this paper is presented in Fig.~\ref{mcALL}. 
For comparison, we show separately the distributions of $|m_{\rm{C}}|$ detections ($\ge3\sigma$),  $|m_{\rm{C}}|$ results $\ge2\sigma$, and all $|m_{\rm{C}}|$ measurements (independently of their significance).

The most evident result with regard to circular polarization is the low level of detection in our observations ($\sim6$\,\%, 8 sources out of 145), which prevents us from conducting a statistical study of circular polarization.
This low detection rate is not surprising given the known low degree of circular polarization of blazars at radio wavelengths \citep[typically $\aplt0.5$\,\% at 2\,cm as observed with VLBI,][]{Homan:2006p238} and the minimum $m_{\rm{C}}$ detection level in our observations ($3\sigma_{m_{\rm{C}}}\apgt0.3$\,\%).

Our circular polarization detection rate is lower than half of that of \citet{Homan:2006p238} ($\sim15$\,\%), which may be explained by the difference in angular resolution between their 15\,GHz VLBA observations and ours.
\citet{Homan:2006p238} showed that most circular polarization is confined to the core of the sources. 
If this also applies to the 86\,GHz emitting regions, our single dish observations (to which all such emitting regions contribute) must reflect lower $m_{\rm{C}}$ and a lower detection rate for our similar sensitivity to circular polarization ($\sigma_{m_{\rm{C}}}\apgt0.1$\,\%).

Among our 8 sources with detected circular polarization, only 3C~84 (0316+413) and 3C~454.3 (2251+158) were also detected at $\apgt3\sigma$ by \citet{Homan:2006p238}.
They found significant circular polarization in the core and in 3 downstream jet emission regions in 3C~84, and in one jet region in 3C~454.3.
The amount and sign of $m_{\rm{C}}$ measured in 2003 by \citet{Homan:2006p238} in the core of 3C~84 is not consistent with our single dish measurement in 2005.
Note though that the VLBI measurement by \citet{Homan:2006p238}, who detected strong circular polarization with opposite sign than in the core within a nearby jet region, may not be comparable with our single dish measurement. In contrast, both $m_{\rm{C}}$ and its sign in the core of 3C~454.3 as measured by Homan \& Lister's, and our measurements are consistent, despite the extreme mm flare displayed by this source during our observations \citep[e.g.,][]{Raiteri:2008p123,Jorstad:subm}.
These discrepancies, as well as the fact that our circular polarization detections do not agree in general with those by \citet{Homan:2006p238}, is consistent with the known circular polarization variability in radio loud AGN \citep[e.g.,][]{Aller:2003p4756}, which may differ considerably at different observing frequencies \citep[e.g.,][]{Homan:2009p6162}.

Circular polarization in AGN jets may be generated either by intrinsically circularly polarized synchrotron radiation, or by Faraday conversion of linear into circular polarization -- see \citet{Wardle:2003p8612} and \citet{Homan:2009p6162} for detailed reviews of the different magnetic field configurations and conditions suitable for both mechanisms.
Both mechanisms have a dependence with the emitting wavelength ($\lambda$), that conspires to reduce $m_{\rm{C}}$ with decreasing $\lambda$, hence providing an additional explanation for the low level of $m_{\rm{C}}$ detection in our observations.
However, whereas for the intrinsic mechanism $m{_{\rm{C}}}{^{\rm{int}}}\propto\lambda^{1/2}$, for Faraday conversion the dependence is much more stronger, $m{_{\rm{C}}}{^{\rm{F}}}\propto\lambda^{5}$ under the assumptions adopted by \citet{Wardle:2003p8612}. 
Hence, whereas $m{_{\rm{C}}}{^{\rm{int}}}$ at 3\,mm is only $\sim2$ times lower than at $2$\,cm, $m{_{\rm{C}}}{^{\rm{F}}}$ is reduced by a factor $\sim6,000$.
Additionally, some of the Faraday conversion scenarios invoked in the literature \citep{Wardle:2003p8612,Ruszkowski:2002p8711,2002A&A...388.1106B,Homan:2009p6162} involve Faraday rotation (internal to the jet), that has a $\Delta\chi^{\rm{F}}\propto\lambda^{2}$ dependence.
All of these considerations render the Faraday conversion mechanism less suitable to explain the circular polarization detected at 3\,mm.
Otherwise it would be difficult to explain that the amount of circular polarization detected in our observations, $0.3{\,\%}\aplt |m{_{\rm{C}}}|\aplt0.7{\,\%}$, is similar to that detected in the MOJAVE survey at 2\,cm \citep{Homan:2006p238}.

\subsection{Influence of Source Variability}
\label{Var}

It is well known that radio loud AGN display in the mm range (as in other spectral ranges) abrupt total flux density variability by up to maxima of $\sim1$ order of magnitude and time scales from months \citep{Jorstad:2005p264, Terasranta:2005p9128, Fuhrmann:2008p267} to days \citep{Agudo:2006p203}.
Such variability is usually connected with the ejection of superluminal emission features from the innermost jet regions \citep[e.g.,][]{Jorstad:2005p264, Kadler:2008p397, Perucho:2008p298}.
The Doppler factor of the source affects its variability time scale in the observer's frame through a directly proportional relation, whereas the cosmological redshift slows down such time scale by a factor $(1+z)^{-1}$.

\citet{Jorstad:2005p264} made a study of the (7, 3, and 1) millimeter polarization variability of 15 radio loud AGN monitored during $\sim3$\,yr.
They identify different classes of variable sources, with the most variable one displaying excursions of up to $\sim1$ order of magnitude in linear polarization degree, and $>90^{\circ}$ in linear polarization angle in time scales of months or even weeks \citep[see also ][for a case of mm and optical polarimetric correlation in {3C~454.3}]{Jorstad:subm}.
At 7\,mm, shorter $m_{\rm{L}}$ and $\chi$ variability time scales of less than a week -- but for smaller variability amplitudes -- are reported by \citet{2007ApJ...659L.107D, Darcangelo:2009p6953}, for the case of the blazars {PKS~0420$-$014}, and {OJ~287}.

However, not all radio loud AGN vary as much as in these extreme cases. 
There seems to be a gradient of variability amplitude and time scales, which is likely related to the intrinsic properties of every source.
Anyhow, source variability is an unavoidable effect that influences our results, as well as those from any other survey at which AGN display significant variability.
Such influence on the selection of our sample and its completeness was discussed in \S~\ref{Intr}.

For the large number of objects in different variability states, our distributions of $S_{86}$, $m_{\rm{L}}$, and $\chi$ are expected to broaden their shapes or to decrease their correlation between them and with other variables, hence conspiring to hidden possible significant results from them. 
However, for such a large sample like ours, we do not expect that variability can fake the statistical results by shifting the distributions towards a particular trend.
From that point of view, we are confident on the results in previous sections, which significance and extension to other subsamples could perhaps have been more ample without the influence of variability.
This is extensive to the studies of $\alpha_{\rm{15,86}}$ and $m_{\rm{L}}/m_{\rm{L,15}}$ in \S~\ref{alpha} and \ref{Misal}, respectively.
The extra variability at 15 GHz that we could not account for introduces extra widening of the distributions against significant statistical results. 
Even in that case, we were able to obtain some significant results.

\section{Conclusions}
\label{Conc}

We have performed the first large 3.5\,mm polarization survey of radio loud AGN over a sample of 145 bright and flat radio spectrum sources, dominated by blazars, observed with the IRAM 30\,m Telescope.
At such wavelengths, Faraday rotation and depolarization are expected to have a weak effect on the observed linear polarization, which thus reflects the intrinsic linear polarization properties of the sources with high fidelity. 

We detect linear and circular polarization above $3\sigma$ levels for 76\,\% and 6\,\% of the sample, respectively.
We have shown that the fractional linear polarization at 86\,GHz clearly exceeds that at 15\,GHz by a factor of $\sim2$ for all classes of sources considered here.
This implies both a larger degree of magnetic field order in the region where the bulk of the 86\,GHz emission is produced, and considerable Faraday depolarization at 15\,GHz, although we can not quantify the relative magnitude of these effects with the present data.

Consistent with previous work suggesting that the class of  $\gamma$-ray bright blazars possess, in general, larger Doppler factors than weaker  $\gamma$-ray bright blazars, we have found that LBAS quasars from our sample are more luminous than non-LBAS quasars.
However, our data do not allow us to claim differences in the luminosity distributions of LBAS and non-LBAS BL~Lac objects.

Our entire source sample shows a trend of lower total flux luminosity for larger degrees of linear polarization, indicating that the level of magnetic field order in the innermost jet emitting regions is anti-correlated with jet luminosity.

Unlike other authors, who observed at radio frequencies (hence affected by Faraday rotation), we do not find a clear relation between the linear polarization angle and the jet structural position angle of BL~Lacs.
This is also true for the remainder of the source subsamples considered in this work, despite the tendency of the electric-vector linear polarization angle to align with the jet position angle found at radio frequencies on quasars.
This implies that the sources in our sample do not contain the conditions required for their linear polarization angles to lie either parallel or perpendicular to the jet axis (in the observer's frame), i.e., that they have a markedly non axisymmetric character.
Indeed, the 3-dimensional character of a non-negligible population of jets in AGN at different scales is known to be complex \cite[e.g.,][]{Fomalont:2000p4880,2003ApJ...589..733P,Jorstad:2005p264,Lister:2005p261,Agudo:2006p331, Agudo:2007p132}.

In the case of purely axisymmetric jets, cancellation of linearly polarized emission parallel and perpendicular to the jet axis forces $\chi$ to align only either parallel, or perpendicular to the axis \citep[e.g,][]{Lyutikov:2005p321,Cawthorne:2006p409}.
Our results contradict this phenomenology, which suggests that the theoretical requirements for relativistic jets to show such behavior are not satisfied when observed at 3\,mm.
This implies that the 3\,mm emitting region in relativistic jets in blazars is primarily non axially-symmetric at this wavelength.
Hence, either the magnetic field or the emitting particle distributions (or both) responsible for the synchrotron radiation at 3\,mm in a large fraction of blazars must have a markedly 3 dimensional (non-axisimmetric) character in order to account for the $|\chi-\phi_{\rm{jet}}|$ distributions of our observations.

We have shown that the circularly polarized emission detected from our observations is most likely generated by intrinsic synchrotron emission. 
This has an important consequence. 
Intrinsic circular polarization cannot be produced in pure pair (electron-positron, $e^{+}e^{-}$) plasma jets.
Thus, sources in which such mechanisms govern circular polarization production must be composed, at least in part, by an electron-proton ($p^{+}e^{-}$) plasma.
This may be taken as a tool for AGN jet composition diagnostics through future mm wavelength polarimetric surveys.
If a large fraction of  sources show $m_{\rm{C}}$ with moduli as large as those detected at cm wavelengths, this may imply that their mm jet emitting regions are primarily composed of $p^{+}e^{-}$ plasma.

\begin{acknowledgements}
     The authors acknowledge the anonymous referee for his/her constructive revision of this paper, which allowed us to improve it considerably.
     They also thank A.~P. Marscher, J.~L. G\'{o}mez, S.~G. Jorstad, and Y. Mizuno for helpful comments on this work.
     This paper is based on observations carried out with the IRAM 30~m Telescope. 
     The authors acknowledge the observers and technicians involved in the operation of the telescope during our observations. 
     In particular, we are grateful to M. Ruiz and J.~L. Santar\'en.    
     IRAM is supported by INSU/CNRS (France), MPG (Germany), and IGN (Spain). 
     I. A. acknowledges support by an I3P contract with the Spanish ``Consejo Superior de Investigaciones Cient\'{i}ficas". 
     He also acknowledges support by the ``Ministerio de Ciencia e Innovaci\'on" of Spain and by the National Science Foundation of the USA through grants AYA2007-67627-C03-03, and AST-0907893, respectively.
     This research has made use of the NASA/IPAC Extragalactic Database, the MOJAVE database, the one by Blazar Group at the Boston University, as well as the USNO Radio Reference Frame Image Database.
\end{acknowledgements}




\begin{deluxetable}{ccccccccccc}
\tablecolumns{11}
\tabletypesize{\scriptsize}
\tablewidth{0pt} 
\tablecaption{\label{T1} Source properties.}
\tablehead{\colhead{Source name}        &   \colhead{}        &     \colhead{R.A.}     &     \colhead{Dec.}     & 
                  \colhead{}   & \colhead{Opt}    &   \colhead{$V$} & \colhead{$\phi_{\rm{jet}}$} & 
                  \colhead{$\nu_{\rm{obs}}$} & \colhead{Ref.} & \colhead{LBAS}  \\
                  \colhead{(IAU)}             &   \colhead{Alias} & \colhead{(J2000.0)} &  \colhead{(J2000.0)} &  
                  \colhead{z} & \colhead{Cl.}  &  \colhead{mag} & \colhead{[$^{\circ}$]} & \colhead{[GHz]} & 
                  \colhead{($\phi_{\rm{jet}}$)}  & \colhead{?} \\
                  \colhead{(1)} &  \colhead{(2)} &  \colhead{(3)} &  \colhead{(4)} &  \colhead{(5)} &  \colhead{(6)} &  
                  \colhead{(7)} &  \colhead{(8)}  & \colhead{(9)} &  \colhead{(10)}  & \colhead{(11)}}
\startdata
{0003+380}            &       \nodata & 00 05 57.1352 & +38 20 14.869 & 0.229 & G & 19.9 & 106 &  15 &   9 & \nodata \\ 
{0048-097}            &       \nodata & 00 50 41.3193 & -09 29 05.122 &  \nodata  & B & 17.4 &   7 &  86 &   6 &  Y  \\ 
{0059+581}\tablenotemark{a} &       \nodata & 01 02 45.7623 & +58 24 11.136 & 0.644\tablenotemark{b} & Q & 17.3\tablenotemark{c} & 235 &  15 &   8 & \nodata \\ 
{0106+013}            &  {4C~01.02} & 01 08 38.7684 & +01 35 00.421 & 2.107 & Q & 18.4 & 235 &  15 &   8 & \nodata \\ 
{0112-017}            &       \nodata & 01 15 17.0917 & -01 27 04.456 & 1.365 & Q & 17.5 & 118 &  15 &   5 & \nodata \\ 
{0113-118}            &       \nodata & 01 16 12.5176 & -11 36 15.412 & 0.672 & Q & 19.0 & 338 &  15 &   9 & \nodata \\ 
{0119+041}            &       \nodata & 01 21 56.8557 & +04 22 24.842 & 0.637 & Q & 19.5 & 124 &  15 &   5 & \nodata \\ 
{0133+476}            &     {DA~55} & 01 36 58.5910 & +47 51 29.164 & 0.859 & Q & 19.5 & 334 &  86 &   6 &  Y  \\ 
{0135-247}            &       \nodata & 01 37 38.3418 & -24 30 53.698 & 0.835 & Q & 17.3 &  75 &   5 &   3 & \nodata \\ 
{0202+149}            &  {4C~15.05} & 02 04 50.4141 & +15 14 11.214 & 0.405 & Q\tablenotemark{d} & 21.0 & 338 &  15 &   8 & \nodata \\ 
{0212+735}            &       \nodata & 02 17 30.7735 & +73 49 32.845 & 2.367 & Q & 20.0 & 113 &  86 &   6 & \nodata \\ 
{0219+428}            &    {3C~66A} & 02 22 39.6114 & +43 02 07.799 & 0.444 & B & 15.2 & 195 &  43 &   4 &  Y  \\ 
{0221+067}            &       \nodata & 02 24 28.4237 & +06 59 23.499 & 0.511 & Q\tablenotemark{d} & 20.7 & 300 &  15 &   9 & \nodata \\ 
{0224+671}\tablenotemark{a} &  {4C~67.05} & 02 28 50.0655 & +67 21 03.123 & 0.523\tablenotemark{b} & Q & 19.5\tablenotemark{e} &   4 &  15 &   8 & \nodata \\ 
{0234+285}            &  {4C~28.07} & 02 37 52.3845 & +28 48 09.782 & 1.207 & Q & 19.3 & 283 &  15 &   8 &  Y  \\ 
{0235+164}            &       \nodata & 02 38 38.9268 & +16 36 59.287 & 0.940 & B\tablenotemark{d} & 18.0\tablenotemark{e} & 270 &  86 &   6 &  Y  \\ 
{0239+108}\tablenotemark{a} &  {4C~+6.11} & 02 42 29.1773 & +11 01 00.856 & 2.680\tablenotemark{b} & Q & 20.0\tablenotemark{e} & 121 &   5 &   3 & \nodata \\ 
{0300+470}            &       \nodata & 03 03 35.2431 & +47 16 16.387 &  \nodata  & B & 16.9 & 126 &  86 &   6 & \nodata \\ 
{0316+413}            &     {3C~84} & 03 19 48.1540 & +41 30 42.160 & 0.017 & G & 12.5 & 186 &  86 &   6 &  Y  \\ 
{0333+321}            &  {NRAO~140} & 03 36 30.0022 & +32 18 28.762 & 1.259 & Q & 17.5 & 128 &  15 &   8 & \nodata \\ 
{0336-019}            &    {CTA~26} & 03 39 30.9336 & -01 46 35.755 & 0.852 & Q & 18.4 &  74 &  86 &   6 & \nodata \\ 
{0355+508}            &  {NRAO~150} & 03 59 29.7464 & +50 57 50.230 & 1.517\tablenotemark{f} & Q & 22.9 & 161 &  43 &   1 & \nodata \\ 
{0403-132}            &       \nodata & 04 05 33.9795 & -13 08 14.345 & 0.571 & Q & 17.1 & 179 &  15 &   8 & \nodata \\ 
{0415+379}            &    {3C~111} & 04 18 21.2682 & +38 01 35.574 & 0.049 & G & 18.1 &  68 &  86 &   6 & \nodata \\ 
{0420-014}            &       \nodata & 04 23 15.7959 & -01 20 33.124 & 0.915 & Q & 17.0 & 192 &  86 &   6 &  Y  \\ 
{0422+004}            &       \nodata & 04 24 46.8226 & +00 36 08.702 & 0.476 & B & 17.0 &   4 &  15 &   8 & \nodata \\ 
{0430+052}            &    {3C~120} & 04 33 11.0894 & +05 21 15.549 & 0.033 & G & 15.1 & 244 &  86 &   6 & \nodata \\ 
{0439+360}\tablenotemark{a} &       \nodata & 04 42 53.3565 & +36 06 52.668 &  \nodata  & U &  \nodata & \nodata & \nodata & \nodata & \nodata \\ 
{0454-234}            &       \nodata & 04 57 03.1634 & -23 24 52.367 & 1.003 & Q & 18.9 & 205 &  15 &   9 &  Y  \\ 
{0458-020}            & {4C~-02.19} & 05 01 12.8003 & -01 59 13.756 & 2.291 & Q & 18.1 & 312 &  15 &   8 & \nodata \\ 
{0514-161}\tablenotemark{a} &       \nodata & 05 16 15.9268 & -16 03 07.614 & 1.278\tablenotemark{g} & Q & 17.0\tablenotemark{e} & \nodata & \nodata & \nodata & \nodata \\ 
{0528+134}            &       \nodata & 05 30 56.4348 & +13 31 55.173 & 2.070 & Q & 20.0 &  75 &  86 &   6 &  Y  \\ 
{0529+075}\tablenotemark{a} &       \nodata & 05 32 38.9895 & +07 32 43.314 & 1.254\tablenotemark{b} & Q & 19.0\tablenotemark{e} & 322 &  15 &   8 & \nodata \\ 
{0552+398}            &    {DA~193} & 05 55 30.7409 & +39 48 49.125 & 2.363 & Q & 18.3 & 288 &  15 &   8 & \nodata \\ 
{0605-085}            &       \nodata & 06 07 59.6922 & -08 34 49.988 & 0.872 & Q & 17.6 & 132 &  15 &   8 & \nodata \\ 
{0607-157}            &       \nodata & 06 09 40.9611 & -15 42 40.476 & 0.324 & Q & 18.0 & 189 &  86 &   6 & \nodata \\ 
{0642+449}            &       \nodata & 06 46 32.0222 & +44 51 16.585 & 3.408 & Q & 18.5 &  93 &  86 &   6 & \nodata \\ 
{0716+714}            &       \nodata & 07 21 53.4701 & +71 20 36.392 & 0.310\tablenotemark{k} & B & 15.5 &  35 &  15 &   8 &  Y  \\ 
{0727-115}            &       \nodata & 07 30 19.1082 & -11 41 12.692 & 1.591 & Q & 20.3 & 254 &  15 &   8 & \nodata \\ 
{0735+178}            &       \nodata & 07 38 07.3910 & +17 42 18.980 & 0.424 & B & 16.2 &  59 &  86 &   6 &  Y  \\ 
{0736+017}            &       \nodata & 07 39 18.0300 & +01 37 04.580 & 0.191 & Q & 16.5 & 275 &  86 &   6 & \nodata \\ 
{0745+241}            &       \nodata & 07 48 36.1316 & +24 00 23.988 & 0.409 & Q\tablenotemark{i} & 19.6 & 296 &  15 &   5 & \nodata \\ 
{0754+100}            &       \nodata & 07 57 06.6602 & +09 56 34.658 & 0.266 & B & 15.0 &  15 &  15 &   8 & \nodata \\ 
{0804+499}            &       \nodata & 08 08 39.6704 & +49 50 36.481 & 1.432 & Q & 19.2 & 129 &  15 &   8 & \nodata \\ 
{0805-077}            &       \nodata & 08 08 15.5274 & -07 51 10.051 & 1.837 & Q & 19.8 & 341 &  15 &   8 & \nodata \\ 
{0814+425}            &    {OJ~425} & 08 18 16.0034 & +42 22 45.337 & 0.245\tablenotemark{j} & B & 18.2 &  91 &  15 &   8 &  Y  \\ 
{0820+560}            &       \nodata & 08 24 47.2441 & +55 52 42.585 & 1.417 & Q & 18.2 &  58 &   5 &   3 &  Y  \\ 
{0823+033}            &       \nodata & 08 25 50.3546 & +03 09 24.408 & 0.506 & B & 16.8 &  87 &  86 &   6 & \nodata \\ 
{0827+243}            &       \nodata & 08 30 52.0861 & +24 10 59.821 & 0.939 & Q & 17.3 & 127 &  15 &   8 & \nodata \\ 
{0829+046}            &       \nodata & 08 31 48.8769 & +04 29 39.086 & 0.180 & B & 16.4 &  67 &  15 &   8 & \nodata \\ 
{0834-201}            &       \nodata & 08 36 39.2094 & -20 16 59.530 & 2.752 & Q & 18.5 & 233 &  15 &   9 & \nodata \\ 
{0836+710}            &  {4C~71.07} & 08 41 24.3819 & +70 53 41.760 & 2.218 & Q & 17.3 & 221 &  15 &   8 & \nodata \\ 
{0851+202}            &    {OJ~287} & 08 54 48.8748 & +20 06 30.572 & 0.306 & B & 15.4 & 241 &  86 &   6 &  Y  \\ 
{0923+392}            & {4C~+39.25} & 09 27 03.0102 & +39 02 20.692 & 0.698 & Q & 17.0 &  88 &  15 &   8 & \nodata \\ 
{0945+408}            &  {4C~40.24} & 09 48 55.3341 & +40 39 44.446 & 1.252 & Q & 18.1 & 159 &  86 &   6 & \nodata \\ 
{0953+254}            &       \nodata & 09 56 49.8762 & +25 15 15.901 & 0.712 & Q & 17.2 & 116 &  15 &   5 & \nodata \\ 
{0954+658}            &       \nodata & 09 58 47.2617 & +65 33 54.666 & 0.367 & B & 16.8 & 289 &  86 &   6 & \nodata \\ 
{1012+232}            &       \nodata & 10 14 47.0622 & +23 01 16.454 & 0.565 & Q & 17.8 & 109 &  86 &   6 & \nodata \\ 
{1034-293}            &       \nodata & 10 37 16.0817 & -29 34 02.914 & 0.312 & Q & 16.5 &  34 &   5 &   3 & \nodata \\ 
{1039+811}            &       \nodata & 10 44 23.1009 & +80 54 39.319 & 1.260 & Q & 17.9 & 300 &   5 &   3 & \nodata \\ 
{1044+719}            &       \nodata & 10 48 27.6375 & +71 43 35.788 & 1.150 & Q & 19.0 & 178 &   8 &  10 & \nodata \\ 
{1045-188}            &       \nodata & 10 48 06.6157 & -19 09 35.965 & 0.595 & Q & 18.2 & 150 &  15 &   8 & \nodata \\ 
{1055+018}            & {4C~+01.28} & 10 58 29.5968 & +01 33 58.860 & 0.888 & Q & 18.3 & 308 &  15 &   8 &  Y  \\ 
{1116+128}            & {4C~+12.39} & 11 18 57.2988 & +12 34 41.549 & 2.118 & Q & 19.2 &  17 &   5 &   3 & \nodata \\ 
{1124-186}            &       \nodata & 11 27 04.3922 & -18 57 17.712 & 1.048 & Q & 18.6 & 169 &  15 &   8 & \nodata \\ 
{1127-145}            &       \nodata & 11 30 07.0525 & -14 49 27.387 & 1.187 & Q & 16.9 &  98 &  15 &   8 &  Y  \\ 
{1144+402}            &       \nodata & 11 46 58.2966 & +39 58 34.085 & 1.089 & Q & 18.0 &   5 &   8 &  10 & \nodata \\ 
{1156+295}            &  {4C~29.45} & 11 59 31.8339 & +29 14 43.608 & 0.729 & Q & 14.4 &  38 &  86 &   6 &  Y  \\ 
{1213-172}\tablenotemark{a} &       \nodata & 12 15 46.6892 & -17 31 45.583 &  \nodata  & U & 21.4\tablenotemark{e} & 104 &  15 &   8 & \nodata \\ 
{1226+023}            &    {3C~273} & 12 29 06.6971 & +02 03 08.453 & 0.158 & Q & 12.8 & 208 &  15 &   8 &  Y  \\ 
{1228+126}            &      {M~87} & 12 30 49.4233 & +12 23 28.043 & 0.004 & G & 12.9 & 262 &  15 &   8 & \nodata \\ 
{1244-255}            &       \nodata & 12 46 46.7983 & -25 47 49.292 & 0.638 & Q & 17.4 & 138 &  15 &   9 &  Y  \\ 
{1253-055}            &    {3C~279} & 12 56 11.1688 & -05 47 21.695 & 0.538 & Q & 17.8 & 215 &  15 &   8 &  Y  \\ 
{1308+326}            &       \nodata & 13 10 28.6573 & +32 20 43.621 & 0.997 & Q & 15.2 & 281 &  86 &   6 &  Y  \\ 
{1328+307}            &    {3C~286} & 13 31 08.2880 & +30 30 32.966 & 0.846 & Q & 17.2 & 230 &  15 &   9 & \nodata \\ 
{1334-127}            &       \nodata & 13 37 39.7841 & -12 57 24.868 & 0.539 & Q & 19.0 & 166 &  15 &   8 & \nodata \\ 
{1354+195}            &    {DA~354} & 13 57 04.4305 & +19 19 07.251 & 0.719 & Q & 16.0 & 146 &  15 &   9 & \nodata \\ 
{1406-076}            &       \nodata & 14 08 56.4811 & -07 52 26.665 & 1.493 & Q & 19.6 & 265 &  43 &   2 & \nodata \\ 
{1413+135}            &       \nodata & 14 15 58.8108 & +13 20 23.601 & 0.247 & B & 20.5 & 247 &  15 &   8 & \nodata \\ 
{1418+546}            &       \nodata & 14 19 46.5784 & +54 23 14.616 & 0.152 & B & 15.7 & 133 &  15 &   9 & \nodata \\ 
{1502+106}            &       \nodata & 15 04 24.9752 & +10 29 39.080 & 1.839 & Q & 18.6 & 131 &  86 &   6 &  Y  \\ 
{1504-166}            &       \nodata & 15 07 04.7876 & -16 52 30.238 & 0.876 & Q & 18.5 & 205 &  15 &   8 & \nodata \\ 
{1510-089}            &       \nodata & 15 12 50.5321 & -09 05 59.845 & 0.360 & Q & 16.5 & 353 &  86 &   6 &  Y  \\ 
{1514-241}            &    {AP~Lib} & 15 17 41.8190 & -24 22 19.431 & 0.048 & B & 14.8 & 173 &  15 &   9 &  Y  \\ 
{1546+027}            &       \nodata & 15 49 29.4326 & +02 37 01.069 & 0.412 & Q & 17.4 & 183 &  86 &   6 & \nodata \\ 
{1548+056}            &       \nodata & 15 50 35.2658 & +05 27 10.400 & 1.422 & Q & 19.5 &   1 &  86 &   6 & \nodata \\ 
{1606+106}            &       \nodata & 16 08 46.1974 & +10 29 07.666 & 1.226 & Q & 18.7 & 336 &  15 &   8 & \nodata \\ 
{1611+343}            &       \nodata & 16 13 41.0330 & +34 12 47.707 & 1.401 & Q & 18.1 & 164 &  15 &   8 & \nodata \\ 
{1622-297}            &       \nodata & 16 26 06.0237 & -29 51 26.770 & 0.815 & Q & 19.5 & 252 &  43 &   2 & \nodata \\ 
{1633+382}            &  {4C~38.41} & 16 35 15.4848 & +38 08 04.423 & 1.807 & Q & 18.0 & 288 &  15 &   8 &  Y  \\ 
{1637+574}            &       \nodata & 16 38 13.4457 & +57 20 23.874 & 0.751 & Q & 16.9 & 225 &  86 &   6 & \nodata \\ 
{1638+398}            &  {NRAO~512} & 16 40 29.6235 & +39 46 45.979 & 1.666 & Q & 19.4 & 282 &  15 &   8 & \nodata \\ 
{1641+399}            &    {3C~345} & 16 42 58.8001 & +39 48 36.958 & 0.594 & Q & 16.6 & 275 &  15 &   8 & \nodata \\ 
{1642+690}            &       \nodata & 16 42 07.8336 & +68 56 39.698 & 0.751 & Q\tablenotemark{i} & 20.5 & 164 &  15 &   5 & \nodata \\ 
{1652+398}            &   {Mrk~501} & 16 53 52.2167 & +39 45 36.609 & 0.033 & B & 13.8 & 125 &  15 &   5 &  Y  \\ 
{1655+077}            &       \nodata & 16 58 09.0340 & +07 41 26.852 & 0.621 & Q & 20.0 & 333 &  86 &   6 & \nodata \\ 
{1657-261}\tablenotemark{a} &       \nodata & 17 00 53.1591 & -26 10 51.478 &  \nodata  & U &  \nodata & \nodata & \nodata & \nodata & \nodata \\ 
{1716+686}            &       \nodata & 17 16 13.9209 & +68 36 38.684 & 0.777 & Q & 18.5 & 326 &   5 &  11 & \nodata \\ 
{1730-130}             &  {NRAO~530} & 17 33 02.7019 & -13 04 49.502 & 0.902 & Q & 19.5 &  31 &  15 &   8 & \nodata \\ 
{1732+389}            &       \nodata & 17 34 20.5664 & +38 57 51.398 & 0.976 & Q & 20.6 &  89 &   5 &   3 & \nodata \\ 
{1739+522}            &       \nodata & 17 40 36.9634 & +52 11 43.410 & 1.379 & Q & 18.7 &  41 &  86 &   6 & \nodata \\ 
{1741-038}             &       \nodata & 17 43 58.8510 & -03 50 04.604 & 1.057 & Q & 20.4 & 237 &  86 &   6 & \nodata \\ 
{1749+096}            &    {OT~081} & 17 51 32.8104 & +09 39 00.700 & 0.320 & B\tablenotemark{k} & 16.8 &   6 &  15 &   8 &  Y  \\ 
{1800+440}            &       \nodata & 18 01 32.2950 & +44 04 21.849 & 0.663 & Q & 17.9 & 271 &  86 &   6 & \nodata \\ 
{1803+784}            &       \nodata & 18 00 45.6222 & +78 28 04.022 & 0.680 & B\tablenotemark{k} & 15.9 & 224 &  86 &   6 &  Y  \\ 
{1807+698}            &    {3C~371} & 18 06 50.6518 & +69 49 28.089 & 0.050 & B & 14.2 & 264 &  15 &   5 & \nodata \\ 
{1823+568}\tablenotemark{a} &  {4C~56.27} & 18 24 07.0480 & +56 51 01.484 & 0.664\tablenotemark{l} & B & 18.4\tablenotemark{c} & 194 &  86 &   6 & \nodata \\ 
{1828+487}            &    {3C~380} & 18 29 31.8047 & +48 44 46.496 & 0.692 & Q & 16.8 & 316 &  86 &   6 & \nodata \\ 
{1830-211}            &       \nodata & 18 33 39.9093 & -21 03 40.049 & 2.507 & Q & 18.7 & \nodata & \nodata & \nodata & \nodata \\ 
{1842+681}            &       \nodata & 18 42 33.7085 & +68 09 25.034 & 0.475 & Q & 18.1 & 117 &  86 &   6 & \nodata \\ 
{1908-201}            &       \nodata & 19 11 09.6517 & -20 06 54.989 & 1.119 & Q & 18.1 &   4 &  15 &   9 &  Y  \\ 
{1921-293}            &       \nodata & 19 24 51.0545 & -29 14 29.838 & 0.352 & Q & 18.2 & 335 &  86 &   6 & \nodata \\ 
{1923+210}\tablenotemark{a} &       \nodata & 19 25 59.5932 & +21 06 26.106 &  \nodata  & U & 16.1\tablenotemark{c} & 245 &  86 &   6 & \nodata \\ 
{1928+738}            &       \nodata & 19 27 48.4595 & +73 58 01.592 & 0.303 & Q & 16.1 & 156 &  15 &   8 & \nodata \\ 
{1954+513}            &       \nodata & 19 55 42.7230 & +51 31 48.585 & 1.223 & Q & 18.5 & 306 & 43 &   7 & \nodata \\ 
{1957+405}            &     {Cyg~A} & 19 59 28.3546 & +40 44 02.101 & 0.056 & G & 15.1 & 283 &  15 &   8 & \nodata \\ 
{1958-179}            &       \nodata & 20 00 57.0848 & -17 48 57.547 & 0.652 & Q & 18.6 & 207 &  15 &   8 & \nodata \\ 
{2005+403}            &       \nodata & 20 07 44.9340 & +40 29 48.622 & 1.736 & Q & 19.0 & 120 &  15 &   8 & \nodata \\ 
{2007+777}            &       \nodata & 20 05 30.9646 & +77 52 43.294 & 0.342 & B & 16.7 & 282 &  86 &   6 & \nodata \\ 
{2013+370}            &       \nodata & 20 15 28.7151 & +37 10 59.640 &  \nodata  & B\tablenotemark{m} & 21.6\tablenotemark{c} & 176 &  86 &   6 & \nodata \\ 
{2021+317}\tablenotemark{a} & {4C~+31.56} & 20 23 19.0066 & +31 53 02.395 &  \nodata  & U & 19.0\tablenotemark{c} & 168 &  15 &   8 & \nodata \\ 
{2023+336}            &       \nodata & 20 25 10.8256 & +33 43 00.265 & 0.219 & B &  \nodata & 344 &  86 &   6 & \nodata \\ 
{2037+511}            &    {3C~418} & 20 38 37.0188 & +51 19 12.687 & 1.687 & Q & 21.0 & 226 &  86 &   6 & \nodata \\ 
{2059+034}            &       \nodata & 21 01 38.8275 & +03 41 31.381 & 1.015 & Q & 17.8 &  30 &   8 &  10 & \nodata \\ 
{2113+293}            &       \nodata & 21 15 29.3850 & +29 33 38.540 & 1.514 & Q & 20.6 & 177 &  15 &   5 & \nodata \\ 
{2121+053}            &       \nodata & 21 23 44.4941 & +05 35 22.192 & 1.941 & Q & 20.4 & 276 &  15 &   8 & \nodata \\ 
{2128-123}            &       \nodata & 21 31 35.2540 & -12 07 04.725 & 0.501 & Q & 16.1 & 216 &  15 &   8 & \nodata \\ 
{2131-021}            &       \nodata & 21 34 10.3053 & -01 53 17.163 & 1.284 & B\tablenotemark{k} & 19.0 & 100 &  15 &   8 & \nodata \\ 
{2134+004}            &    {DA~553} & 21 36 38.5791 & +00 41 54.319 & 1.932 & Q & 17.1 & 332 &  15 &   8 & \nodata \\ 
{2136+141}            &       \nodata & 21 39 01.3021 & +14 23 36.108 & 2.427 & Q & 18.9 & 310 &  15 &   8 & \nodata \\ 
{2145+067}            & {4C~+06.69} & 21 48 05.4509 & +06 57 38.710 & 0.999 & Q & 16.5 & 122 &  15 &   8 & \nodata \\ 
{2155-152}            &       \nodata & 21 58 06.2819 & -15 01 09.327 & 0.672 & Q & 18.3 & 207 &  15 &   8 & \nodata \\ 
{2200+420}            &    {BL~Lac} & 22 02 43.2793 & +42 16 40.073 & 0.069 & B & 14.7 & 221 &  86 &   6 &  Y  \\ 
{2201+315}            &  {4C~31.63} & 22 03 14.9665 & +31 45 38.359 & 0.298 & Q & 15.6 & 209 &  15 &   8 & \nodata \\ 
{2210-257}            &       \nodata & 22 13 02.4963 & -25 29 30.054 & 1.831 & Q & 19.0 &  91 &   5 &   3 & \nodata \\ 
{2216-038}            &       \nodata & 22 18 52.0315 & -03 35 36.837 & 0.901 & Q & 16.4 & 188 &  15 &   8 & \nodata \\ 
{2223-052}            &    {3C~446} & 22 25 47.2570 & -04 57 01.271 & 1.404 & Q & 18.4 & 128 &  86 &   6 & \nodata \\ 
{2230+114}            &   {CTA~102} & 22 32 36.4015 & +11 43 50.985 & 1.037 & Q & 17.3 & 126 &  15 &   8 &  Y  \\ 
{2234+282}            &       \nodata & 22 36 22.4627 & +28 28 57.525 & 0.795 & Q & 19.0 & 263 &  15 &   5 & \nodata \\ 
{2243-123}            &       \nodata & 22 46 18.2309 & -12 06 51.110 & 0.630 & Q & 16.4 & 359 &  15 &   8 & \nodata \\ 
{2251+158}            &  {3C~454.3} & 22 53 57.7438 & +16 08 53.648 & 0.859 & Q & 16.1 & 255 &  86 &   6 &  Y  \\ 
{2254+617}\tablenotemark{a} &       \nodata & 22 56 17.9320 & +62 01 49.545 &  \nodata  & U &  \nodata & \nodata & \nodata & \nodata & \nodata \\ 
{2255-282}            &       \nodata & 22 58 05.9656 & -27 58 21.312 & 0.927 & Q & 16.8 & 224 &  15 &   9 & \nodata \\ 
{2318+049}            &       \nodata & 23 20 44.8503 & +05 13 50.085 & 0.623 & Q & 19.0 & 322 &  15 &   9 & \nodata \\ 
{2345-167}            &       \nodata & 23 48 02.6085 & -16 31 12.022 & 0.576 & Q & 18.4 & 121 &  86 &   6 & \nodata \\ 
\enddata
\tablecomments{Columns indicate, for each line: (1) IAU B1950.0 source name, (2) common source name, 
                         (3) and (4) J2000.0 observing right ascension and declination, respectively, 
                         (5) redshift from \citet{VeronCetty:2006p4900} catalog, 
                         (6) optical classification from  \citet{VeronCetty:2006p4900} catalog, 
                         (7) V magnitude from  \citet{VeronCetty:2006p4900} catalog, 
                         (8) jet position angle ($\phi_{\rm{jet}}$), 
                         (9) VLBI observing frequency of the image from which $\phi_{\rm{jet}}$ was measured, 
                         (10) references from which $\phi_{\rm{jet}}$ was obtained,                           
                         (11) source in the LBAS list of sources (Y), or not (...).}
\tablerefs{(1) \citet{Agudo:2007p132}; (2) BU- Blazar Group Web Page; (3) \citet{Fomalont:2000p4880}; 
                (4) \citet{Jorstad:2005p264}; (5) \citet{Kellermann:2004p3406}; (6) \citet{Lee:2008p301}; 
                (7) \citet{Lister:2001p994}; (8) \citet{Lister:2005p261}; (9) MOJAVE Web Page; 
                (10) USNO Radio Reference Frame Image Database Web Page; 
                (11) \citet{Xu:1995p9669}.}
\tablenotetext{a}{Not present in \citet{VeronCetty:2006p4900} catalog.}
\tablenotetext{b}{Redshift from \citet{SowardsEmmerd:2005p4877}.}
\tablenotetext{c}{$V$-mag$=R$-mag from NED.}
\tablenotetext{d}{Classified as a possible galaxy in \citet{VeronCetty:2006p4900} catalog.}
\tablenotetext{e}{$V$-mag$=$optical-mag from NED.}
\tablenotetext{f}{From \citet{Agudo:2007p132, AcostaPulido:2010p11611}.}
\tablenotetext{g}{Redshift from \citet{Wills:1976p7176}.}
\tablenotetext{h}{Redshift from \citet{Nilsson:2008p7197}.}
\tablenotetext{i}{Classified as a galaxy in \citet{VeronCetty:2006p4900} catalog.}
\tablenotetext{j}{$z=0.245$ from \citet{Britzen:2008p11942}. \citet{Sbarufatti:2005p7169} give $z>0.75$.}
\tablenotetext{k}{Classified as a quasar in \citet{VeronCetty:2006p4900} catalog.}
\tablenotetext{l}{Redshift from \cite{Lawrence:1986p7166}.}
\tablenotetext{m}{Classified as a possible BL~Lac in \citet{VeronCetty:2006p4900} catalog.}
\end{deluxetable}

\begin{deluxetable}{ccccccc}
\tablecolumns{7}
\tabletypesize{\scriptsize} 
\tablewidth{0pt}  
\tablecaption{\label{T2} Summary of observing results.}
\tablehead{\colhead{Source} & \colhead{}  & \colhead{$t_{\rm{int}}$} & \colhead{$S_{86}$} & 
                  \colhead{$m_{\rm{L}}$} & \colhead{$\chi$} & \colhead{$m_{\rm{C}}$} \\
                  \colhead{name}  & \colhead{Epoch} & \colhead{[s]} &  \colhead{[Jy]} & 
                  \colhead{[\%]} & \colhead{[$^{\circ}$]} & \colhead{[\%]} \\
                  \colhead{(1)} &  \colhead{(2)} &  \colhead{(3)} &  \colhead{(4)} &  
                  \colhead{(5)} &  \colhead{(6)} &  \colhead{(7)}}
\startdata
{0003+380} & Jul. 2005 &   4.0 &  1.24$\pm$0.06 &  2.20$\pm$0.54 &  36.8$\pm$ 7.0 &     $<$0.67    \\ 
{0048-097} & Jul. 2005 &   8.0 &  0.62$\pm$0.03 &     $<$ 1.72    &      \nodata       &     $<$0.89    \\ 
{0059+581} & Sep. 2007 &   8.0 &  1.10$\pm$0.05 &     $<$ 1.58    &      \nodata       &     $<$0.73    \\ 
{0106+013} & Jun. 2009 &   4.0 &  1.71$\pm$0.09 &     $<$ 1.58    &      \nodata       &     $<$0.63    \\ 
{0112-017} & Jun. 2009 &  16.0 &  0.15$\pm$0.01 &     $<$ 4.29    &      \nodata       &     $<$3.79    \\ 
{0113-118} & Jul. 2005 &   8.0 &  0.96$\pm$0.05 &  4.91$\pm$0.53 & 157.0$\pm$ 3.1 &     $<$0.60    \\ 
{0119+041} & Jul. 2005 &   8.0 &  0.42$\pm$0.02 &     $<$ 2.00    &      \nodata       &     $<$1.41    \\ 
{0133+476} & Jul. 2005 &   4.0 &  3.75$\pm$0.19 &  4.11$\pm$0.51 &  41.6$\pm$ 3.6 &     $<$0.44    \\ 
{0135-247} & Jun. 2009 &  12.0 &  1.36$\pm$0.07 &  3.70$\pm$0.55 & 112.1$\pm$ 4.2 &     $<$0.68    \\ 
{0202+149} & Jul. 2005 &  24.0 &  0.42$\pm$0.02 &  5.14$\pm$0.56 & 114.1$\pm$ 3.1 &     $<$0.79    \\ 
{0212+735} & Jul. 2005 &  12.0 &  1.51$\pm$0.08 &  3.94$\pm$0.51 & 144.0$\pm$ 3.7 &     $<$0.45    \\ 
{0219+428} & Nov. 2008 &  16.0 &  0.60$\pm$0.03 &  4.36$\pm$0.54 &   2.6$\pm$ 3.6 &     $<$0.73    \\ 
{0221+067} & Jul. 2005 &  32.0 &  0.54$\pm$0.03 &  6.62$\pm$0.54 &  64.8$\pm$ 2.3 &     $<$0.70    \\ 
{0224+671} & Jul. 2005 &   8.0 &  0.89$\pm$0.04 &  4.29$\pm$0.54 &  22.1$\pm$ 3.6 &     $<$0.66    \\ 
{0234+285} & Jul. 2005 &  16.0 &  3.87$\pm$0.19 &  2.42$\pm$0.50 & 126.7$\pm$ 5.9 &     $<$0.32    \\ 
{0235+164} & Jul. 2005 &  16.0 &  1.96$\pm$0.10 &  4.20$\pm$0.50 &  14.7$\pm$ 3.4 &     $<$0.36    \\ 
{0239+108} & Jul. 2005 &  16.0 &  0.25$\pm$0.01 &  3.78$\pm$0.72 &  27.8$\pm$ 5.4 &     $<$1.59    \\ 
{0300+470} & Jul. 2005 &   8.0 &  0.45$\pm$0.02 &     $<$ 1.83    &      \nodata       &     $<$1.06    \\ 
{0316+413} & Jul. 2005 &   4.0 &  5.95$\pm$0.30 &     $<$ 1.51    &      \nodata       &  0.46$\pm$0.11 \\ 
{0333+321} & Sep. 2007 &   8.0 &  0.79$\pm$0.04 &  1.99$\pm$0.57 &  52.5$\pm$ 8.3 &     $<$0.79    \\ 
{0336-019} & Jul. 2005 &  24.0 &  1.14$\pm$0.06 &  4.20$\pm$0.51 &  84.2$\pm$ 3.5 &     $<$0.43    \\ 
{0355+508} & Jul. 2005 &   4.0 &  4.10$\pm$0.21 &  2.05$\pm$0.50 &  29.3$\pm$ 7.0 &     $<$0.34    \\ 
{0403-132} & Jun. 2009 &  16.0 &  0.38$\pm$0.02 &  4.01$\pm$0.63 & 170.4$\pm$ 4.4 &     $<$1.25    \\ 
{0415+379} & Jul. 2005 &   4.0 &  3.90$\pm$0.19 &     $<$ 1.52    &      \nodata       &     $<$0.37    \\ 
{0420-014} & Jul. 2005 &  12.0 &  2.62$\pm$0.13 &  2.70$\pm$0.50 & 125.6$\pm$ 5.3 &     $<$0.35    \\ 
{0422+004} & Jul. 2005 &  12.0 &  1.60$\pm$0.08 &  5.52$\pm$0.51 & 150.2$\pm$ 2.7 &     $<$0.43    \\ 
{0430+052} & Jul. 2005 &  20.0 &  2.06$\pm$0.10 &     $<$ 1.51    &      \nodata       &     $<$0.35    \\ 
{0439+360} & Jul. 2005 &   4.0 &  1.77$\pm$0.09 &     $<$ 1.55    &      \nodata       &     $<$0.55    \\ 
{0454-234} & Jul. 2005 &   8.0 &  1.19$\pm$0.06 &  4.51$\pm$0.54 &  13.2$\pm$ 3.5 &     $<$0.71    \\ 
{0458-020} & Jul. 2005 &  24.0 &  0.42$\pm$0.02 &  2.78$\pm$0.58 &  89.8$\pm$ 5.9 &     $<$0.87    \\ 
{0514-161} & Feb. 2009 &   8.0 &  0.20$\pm$0.01 &     $<$ 2.95    &      \nodata       &     $<$3.11    \\ 
{0528+134} & Jul. 2005 &  12.0 &  2.00$\pm$0.10 &  2.63$\pm$0.51 &  91.2$\pm$ 5.5 &     $<$0.39    \\ 
{0529+075} & Jul. 2009 &   8.0 &  0.98$\pm$0.05 &  3.84$\pm$0.55 & 149.2$\pm$ 4.0 &     $<$0.66    \\ 
{0552+398} & May  2006 &   8.0 &  1.35$\pm$0.07 &  4.72$\pm$0.54 & 119.6$\pm$ 3.3 &     $<$0.64    \\ 
{0605-085} & Jul. 2005 &   8.0 &  1.03$\pm$0.05 &  4.50$\pm$0.54 &   1.6$\pm$ 3.4 &     $<$0.70    \\ 
{0607-157} & Jul. 2005 &   8.0 &  1.55$\pm$0.08 &  3.12$\pm$0.52 & 138.8$\pm$ 4.8 &     $<$0.53    \\ 
{0642+449} & May  2006 &   8.0 &  1.27$\pm$0.06 &  2.75$\pm$0.54 &   2.6$\pm$ 5.7 &     $<$0.66    \\ 
{0716+714} & Aug. 2007 &   8.0 &  1.89$\pm$0.09 & 11.44$\pm$0.53 & 143.0$\pm$ 1.3 &     $<$0.65    \\ 
{0727-115} & Jul. 2009 &   8.0 &  3.47$\pm$0.17 &  2.55$\pm$0.50 &  99.2$\pm$ 5.7 &     $<$0.35    \\ 
{0735+178} & Feb. 2008 &   8.0 &  0.56$\pm$0.03 &  2.56$\pm$0.61 & 163.5$\pm$ 6.8 &     $<$1.00    \\ 
{0736+017} & Oct. 2009 &  12.0 &  1.68$\pm$0.08 &     $<$ 1.52    &      \nodata       &     $<$0.38    \\ 
{0745+241} & Jul. 2009 &   8.0 &  0.89$\pm$0.04 &  5.72$\pm$0.54 &  99.6$\pm$ 2.7 &     $<$0.62    \\ 
{0754+100} & Jun. 2009 &  16.0 &  1.93$\pm$0.10 &  5.44$\pm$0.51 &  16.6$\pm$ 2.7 &     $<$0.64    \\ 
{0804+499} & Jun. 2009 &  12.0 &  0.44$\pm$0.02 &  2.24$\pm$0.60 & 101.4$\pm$ 7.7 &     $<$1.01    \\ 
{0805-077} & May  2006 &  16.0 &  0.89$\pm$0.04 &  4.71$\pm$0.55 & 175.2$\pm$ 3.4 &     $<$0.68    \\ 
{0814+425} & Jul. 2009 &  16.0 &  1.03$\pm$0.05 &     $<$ 4.76    &      \nodata       &     $<$2.74    \\ 
{0820+560} & Feb. 2009 &   8.0 &  0.50$\pm$0.02 &  2.94$\pm$0.63 &  23.4$\pm$ 6.1 &     $<$1.09    \\ 
{0823+033} & May  2006 &   8.0 &  1.92$\pm$0.10 &  3.38$\pm$0.52 &  27.5$\pm$ 4.4 &  0.60$\pm$0.16 \\ 
{0827+243} & Feb. 2009 &   8.0 &  0.95$\pm$0.05 &     $<$ 1.60    &      \nodata       & -0.63$\pm$0.20 \\ 
{0829+046} & Aug. 2007 &  16.0 &  0.70$\pm$0.04 &  8.20$\pm$0.62 &  45.6$\pm$ 1.9 &     $<$1.21    \\ 
{0834-201} & May  2006 &  16.0 &  1.02$\pm$0.05 &  1.94$\pm$0.55 &  43.8$\pm$ 8.0 &     $<$0.79    \\ 
{0836+710} & May  2006 &   8.0 &  2.53$\pm$0.13 &  5.59$\pm$0.51 & 107.6$\pm$ 2.6 & -0.48$\pm$0.15 \\ 
{0851+202} & Jul. 2005 &   8.0 &  3.20$\pm$0.16 &  6.77$\pm$0.50 & 152.8$\pm$ 2.1 &     $<$0.35    \\ 
{0923+392} & May  2006 &   8.0 &  4.54$\pm$0.23 &  4.16$\pm$0.50 & 137.2$\pm$ 3.5 &     $<$0.34    \\ 
{0945+408} & May  2006 &   8.0 &  0.60$\pm$0.03 &  2.46$\pm$0.67 &  47.0$\pm$ 7.6 &     $<$1.48    \\ 
{0953+254} & Jul. 2005 &   8.0 &  0.90$\pm$0.05 &  1.86$\pm$0.54 &  41.8$\pm$ 8.3 &     $<$0.65    \\ 
{0954+658} & May  2006 &   4.0 &  2.33$\pm$0.12 &  5.20$\pm$0.52 &   8.1$\pm$ 2.9 &     $<$0.82    \\ 
{1012+232} & Jul. 2005 &   8.0 &  0.49$\pm$0.02 &  3.85$\pm$0.61 &  71.7$\pm$ 4.6 &     $<$1.13    \\ 
{1034-293} & Feb. 2009 &   8.0 &  1.20$\pm$0.06 &  2.15$\pm$0.54 & 138.1$\pm$ 7.3 &     $<$0.61    \\ 
{1039+811} & May  2006 &   8.0 &  0.45$\pm$0.02 &     $<$ 2.15    &      \nodata       &     $<$1.57    \\ 
{1044+719} & Jul. 2005 &   8.0 &  1.12$\pm$0.06 &     $<$ 1.61    &      \nodata       &     $<$0.72    \\ 
{1045-188} & Jul. 2005 &   8.0 &  1.05$\pm$0.05 &  6.65$\pm$0.58 & 148.1$\pm$ 2.4 &     $<$0.86    \\ 
{1055+018} & Jul. 2005 &   8.0 &  3.01$\pm$0.15 &  4.01$\pm$0.50 & 132.0$\pm$ 3.6 &     $<$0.35    \\ 
{1116+128} & Feb. 2009 &   8.0 &  0.54$\pm$0.03 &  5.02$\pm$0.63 &  10.7$\pm$ 3.7 &     $<$1.22    \\ 
{1124-186} & Jul. 2005 &   8.0 &  1.27$\pm$0.06 &  9.56$\pm$0.53 &  29.5$\pm$ 1.6 &  0.58$\pm$0.19 \\ 
{1127-145} & Aug. 2007 &   8.0 &  1.31$\pm$0.07 &  2.61$\pm$0.54 &  81.9$\pm$ 6.3 &     $<$0.78    \\ 
{1144+402} & Jul. 2009 &   8.0 &  0.77$\pm$0.04 &  4.87$\pm$0.53 & 117.2$\pm$ 3.1 &     $<$0.58    \\ 
{1156+295} & Aug. 2007 &   8.0 &  0.72$\pm$0.04 &  4.71$\pm$0.58 & 161.4$\pm$ 3.6 &     $<$1.14    \\ 
{1213-172} & Feb. 2009 &   8.0 &  0.93$\pm$0.05 &     $<$ 1.61    &      \nodata       &     $<$0.67    \\ 
{1226+023} & Aug. 2007 &   8.0 & 16.01$\pm$0.80 &  4.14$\pm$0.50 & 131.8$\pm$ 3.5 &     $<$0.31    \\ 
{1228+126} & Jul. 2005 &   4.0 &  3.06$\pm$0.15 &  7.82$\pm$0.52 &  51.4$\pm$ 1.9 &     $<$0.55    \\ 
{1244-255} & Jul. 2005 &   8.0 &  1.66$\pm$0.08 &  2.50$\pm$0.52 & 147.2$\pm$ 5.9 &     $<$0.52    \\ 
{1253-055} & Aug. 2007 &   8.0 & 11.46$\pm$0.57 &     $<$ 1.50    &      \nodata       &     $<$0.31    \\ 
{1308+326} & Jul. 2005 &  24.0 &  0.84$\pm$0.04 &  6.18$\pm$0.54 & 164.9$\pm$ 2.5 &     $<$0.69    \\ 
{1328+307} & Oct. 2006 &  24.0 &  0.97$\pm$0.05 & 13.40$\pm$0.52 &  39.6$\pm$ 1.1 &     $<$0.45    \\ 
{1334-127} & Jul. 2005 &   4.0 &  4.76$\pm$0.24 &  7.13$\pm$0.50 & 172.1$\pm$ 2.0 & -0.35$\pm$0.11 \\ 
{1354+195} & Jul. 2005 &   8.0 &  1.54$\pm$0.08 &  2.38$\pm$0.53 &  21.8$\pm$ 6.4 &     $<$0.58    \\ 
{1406-076} & Aug. 2007 &   8.0 &  0.57$\pm$0.03 &     $<$ 1.93    &      \nodata       &     $<$1.62    \\ 
{1413+135} & Sep. 2009 &  16.0 &  0.31$\pm$0.02 &     $<$ 1.84    &      \nodata       &     $<$1.11    \\ 
{1418+546} & Jul. 2005 &  12.0 &  0.80$\pm$0.04 &  2.80$\pm$0.56 & 107.7$\pm$ 5.7 &     $<$0.85    \\ 
{1502+106} & Jul. 2005 &   8.0 &  1.10$\pm$0.06 &  2.39$\pm$0.60 &  40.1$\pm$ 7.0 &     $<$0.97    \\ 
{1504-166} & Jul. 2005 &   8.0 &  0.71$\pm$0.04 &  2.16$\pm$0.69 &  65.5$\pm$ 9.1 &     $<$1.61    \\ 
{1510-089} & Jul. 2005 &   8.0 &  0.98$\pm$0.05 &  3.93$\pm$0.57 &   5.5$\pm$ 4.2 &     $<$0.90    \\ 
{1514-241} & Jul. 2005 &   8.0 &  1.53$\pm$0.08 &  2.65$\pm$0.52 &  44.7$\pm$ 5.6 &     $<$0.50    \\ 
{1546+027} & Jul. 2005 &  12.0 &  2.48$\pm$0.12 &     $<$ 1.52    &      \nodata       &     $<$0.37    \\ 
{1548+056} & Jul. 2005 &  12.0 &  1.45$\pm$0.07 &  6.49$\pm$0.51 & 153.3$\pm$ 2.3 &     $<$0.49    \\ 
{1606+106} & Jul. 2005 &  16.0 &  0.71$\pm$0.04 &     $<$ 1.65    &      \nodata       &     $<$0.74    \\ 
{1611+343} & Jul. 2005 &   8.0 &  3.42$\pm$0.17 &  1.97$\pm$0.50 & 171.3$\pm$ 7.3 &     $<$0.36    \\ 
{1622-297} & Jul. 2005 &  12.0 &  2.21$\pm$0.11 &  4.56$\pm$0.52 & 178.8$\pm$ 3.2 &     $<$0.48    \\ 
{1633+382} & Jul. 2005 &  12.0 &  2.15$\pm$0.11 &  3.54$\pm$0.51 &  37.7$\pm$ 4.1 &     $<$0.41    \\ 
{1637+574} & Jul. 2005 &  16.0 &  1.46$\pm$0.07 &     $<$ 1.53    &      \nodata       &     $<$0.44    \\ 
{1638+398} & Jul. 2005 &  16.0 &  0.33$\pm$0.02 &     $<$ 2.07    &      \nodata       &     $<$1.45    \\ 
{1641+399} & Jul. 2005 &  12.0 &  3.53$\pm$0.18 &  3.07$\pm$0.50 &  95.9$\pm$ 4.7 &     $<$0.34    \\ 
{1642+690} & Jul. 2005 &  12.0 &  2.54$\pm$0.13 &  5.10$\pm$0.50 &  88.0$\pm$ 2.8 &     $<$0.36    \\ 
{1652+398} & Jul. 2007 &   8.0 &  0.65$\pm$0.03 &  4.42$\pm$0.60 &   3.3$\pm$ 3.9 &     $<$1.05    \\ 
{1655+077} & Jul. 2005 &  16.0 &  0.99$\pm$0.05 &  5.69$\pm$0.52 & 130.6$\pm$ 2.6 &     $<$0.51    \\ 
{1657-261} & Sep. 2009 &  16.0 &  0.69$\pm$0.03 &  5.26$\pm$0.57 & 163.4$\pm$ 3.0 &     $<$0.88    \\ 
{1716+686} & Jul. 2005 &  16.0 &  0.48$\pm$0.02 &     $<$ 1.76    &      \nodata       &     $<$0.96    \\ 
{1730-130} & Jul. 2005 &   8.0 &  1.96$\pm$0.10 &  3.81$\pm$0.51 & 147.7$\pm$ 3.8 &     $<$0.41    \\ 
{1732+389} & Jul. 2005 &  16.0 &  0.74$\pm$0.04 &  5.93$\pm$0.54 & 114.3$\pm$ 2.6 &     $<$0.69    \\ 
{1739+522} & Jul. 2005 &  16.0 &  0.59$\pm$0.03 &  2.08$\pm$0.56 &  99.3$\pm$ 7.8 &     $<$0.84    \\ 
{1741-038} & Jul. 2005 &   4.0 &  3.61$\pm$0.18 &  3.32$\pm$0.50 & 156.4$\pm$ 4.4 &     $<$0.37    \\ 
{1749+096} & Jul. 2005 &   4.0 &  2.79$\pm$0.14 &     $<$ 1.55    &      \nodata       &     $<$0.48    \\ 
{1800+440} & Jul. 2005 &   8.0 &  1.40$\pm$0.07 &  2.51$\pm$0.52 &  81.6$\pm$ 5.9 &     $<$0.53    \\ 
{1803+784} & Jul. 2005 &  16.0 &  1.34$\pm$0.07 &  6.35$\pm$0.51 & 116.9$\pm$ 2.3 &     $<$0.45    \\ 
{1807+698} & Jul. 2005 &  12.0 &  1.12$\pm$0.06 &  2.10$\pm$0.52 &  60.3$\pm$ 7.1 &     $<$0.54    \\ 
{1823+568} & Jul. 2005 &  12.0 &  1.43$\pm$0.07 &  6.95$\pm$0.51 &  16.8$\pm$ 2.1 &     $<$0.45    \\ 
{1828+487} & Jul. 2005 &  12.0 &  2.39$\pm$0.12 &  2.82$\pm$0.50 & 104.5$\pm$ 5.1 &     $<$0.36    \\ 
{1830-211} & Jul. 2005 &   8.0 &  1.76$\pm$0.09 &     $<$ 1.56    &      \nodata       &     $<$0.52    \\ 
{1842+681} & Jul. 2005 &  16.0 &  1.04$\pm$0.05 &  3.34$\pm$0.52 &   3.3$\pm$ 4.5 &     $<$0.49    \\ 
{1908-201} & Jul. 2005 &   8.0 &  3.19$\pm$0.16 &  4.17$\pm$0.51 & 158.8$\pm$ 3.5 &     $<$0.38    \\ 
{1921-293} & Jul. 2005 &   8.0 &  7.68$\pm$0.38 &  2.00$\pm$0.50 & 149.3$\pm$ 7.2 &     $<$0.32    \\ 
{1923+210} & Jul. 2005 &   8.0 &  1.27$\pm$0.06 &     $<$ 1.57    &      \nodata       &     $<$0.57    \\ 
{1928+738} & Jul. 2005 &   8.0 &  1.90$\pm$0.09 &  2.86$\pm$0.51 &  88.6$\pm$ 5.1 &     $<$0.42    \\ 
{1954+513} & Jul. 2005 &   8.0 &  0.70$\pm$0.04 &  7.93$\pm$0.57 & 128.3$\pm$ 2.0 &     $<$0.82    \\ 
{1957+405} & Jul. 2005 &  16.0 &  1.06$\pm$0.05 &     $<$ 1.55    &      \nodata       &     $<$0.46    \\ 
{1958-179} & Jul. 2005 &   8.0 &  1.72$\pm$0.09 &  2.99$\pm$0.52 &  12.0$\pm$ 4.9 &     $<$0.52    \\ 
{2005+403} & Jul. 2005 &  12.0 &  0.90$\pm$0.05 &  2.24$\pm$0.53 & 138.6$\pm$ 6.8 &     $<$0.61    \\ 
{2007+777} & Jul. 2005 &  12.0 &  1.44$\pm$0.07 &  9.20$\pm$0.51 &  90.5$\pm$ 1.6 &     $<$0.46    \\ 
{2013+370} & Jul. 2005 &   8.0 &  1.81$\pm$0.09 &     $<$ 1.53    &      \nodata       &     $<$0.44    \\ 
{2021+317} & Jul. 2005 &   4.0 &  0.49$\pm$0.02 &  4.61$\pm$0.73 & 120.5$\pm$ 4.5 &     $<$1.72    \\ 
{2023+336} & Jul. 2005 &   8.0 &  1.33$\pm$0.07 &  4.17$\pm$0.52 &  51.3$\pm$ 3.6 &     $<$0.56    \\ 
{2037+511} & Jul. 2005 &   8.0 &  2.16$\pm$0.11 &  5.04$\pm$0.51 & 112.6$\pm$ 2.9 &     $<$0.41    \\ 
{2059+034} & Jul. 2005 &   8.0 &  1.10$\pm$0.06 &     $<$ 1.61    &      \nodata       &     $<$0.63    \\ 
{2113+293} & Feb. 2009 &   8.0 &  0.47$\pm$0.02 &  4.23$\pm$0.72 & 164.3$\pm$ 4.8 &     $<$1.51    \\ 
{2121+053} & Jul. 2005 &  32.0 &  1.04$\pm$0.05 &  9.68$\pm$0.51 &  33.6$\pm$ 1.5 &     $<$0.42    \\ 
{2128-123} & Jul. 2005 &   8.0 &  1.42$\pm$0.07 &  3.65$\pm$0.52 & 131.3$\pm$ 4.1 &     $<$0.56    \\ 
{2131-021} & Jul. 2005 &  20.0 &  1.29$\pm$0.06 &  4.71$\pm$0.51 &  76.4$\pm$ 3.1 &     $<$0.40    \\ 
{2134+004} & Jul. 2005 &  12.0 &  1.43$\pm$0.07 &  1.63$\pm$0.51 & 176.3$\pm$ 9.0 &     $<$0.43    \\ 
{2136+141} & Jul. 2005 &  32.0 &  0.65$\pm$0.03 &  4.70$\pm$0.52 & 163.3$\pm$ 3.2 &     $<$0.51    \\ 
{2145+067} & Jul. 2005 &  24.0 &  4.82$\pm$0.24 &  3.55$\pm$0.50 &  36.5$\pm$ 4.0 &     $<$0.31    \\ 
{2155-152} & Sep. 2007 &   8.0 &  1.21$\pm$0.06 & 11.04$\pm$0.53 &  51.9$\pm$ 1.4 &     $<$0.93    \\ 
{2200+420} & Jul. 2005 &  12.0 &  5.97$\pm$0.30 & 10.15$\pm$0.50 & 174.9$\pm$ 1.4 & -0.44$\pm$0.10 \\ 
{2201+315} & Sep. 2007 &   8.0 &  1.66$\pm$0.08 &  4.38$\pm$0.51 & 149.4$\pm$ 3.3 &     $<$0.63    \\ 
{2210-257} & Sep. 2009 &  24.0 &  0.43$\pm$0.02 &  2.08$\pm$0.60 & 148.8$\pm$ 8.2 &     $<$1.03    \\ 
{2216-038} & Jul. 2005 &  12.0 &  1.28$\pm$0.06 &  2.90$\pm$0.52 &  24.6$\pm$ 5.1 &     $<$0.49    \\ 
{2223-052} & Jul. 2005 &   8.0 &  4.18$\pm$0.21 &     $<$ 1.51    &      \nodata       &     $<$0.36    \\ 
{2230+114} & Jul. 2005 &  12.0 &  3.68$\pm$0.18 &     $<$ 1.51    &      \nodata       &     $<$0.33    \\ 
{2234+282} & Jul. 2005 &  12.0 &  0.57$\pm$0.03 &  3.04$\pm$0.58 & 168.6$\pm$ 5.4 &     $<$0.95    \\ 
{2243-123} & Jul. 2005 &   8.0 &  2.42$\pm$0.12 &  3.32$\pm$0.51 & 150.1$\pm$ 4.4 &     $<$0.40    \\ 
{2251+158} & Jul. 2005 &  32.0 & 21.12$\pm$1.06 &  1.85$\pm$0.50 &  48.5$\pm$ 7.8 &  0.36$\pm$0.10 \\ 
{2254+617} & Jul. 2005 &   8.0 &  0.46$\pm$0.03 &     $<$ 1.89    &      \nodata       &     $<$1.21    \\ 
{2255-282} & Jul. 2005 &   4.0 &  1.33$\pm$0.07 &     $<$ 1.65    &      \nodata       &     $<$0.74    \\ 
{2318+049} & Jul. 2005 &  16.0 &  0.73$\pm$0.04 &  4.18$\pm$0.54 & 176.3$\pm$ 3.7 &     $<$0.63    \\ 
{2345-167} & Aug. 2007 &   8.0 &  1.40$\pm$0.07 &  3.53$\pm$0.54 & 148.8$\pm$ 4.6 &     $<$1.00    \\ 
\enddata
\tablecomments{Columns indicate, for each line: (1) IAU B1950.0 source name, 
                         (2) observing epoch, (3) integration time, (4) 86\,GHz flux density, 
                         (5) fractional linear polarization, (6) linear polarization electric vector position angle
                         (7) fractional circular polarization.}
\end{deluxetable}


\end{document}